\documentclass[conference]{IEEEtran} 

\usepackage[dvipsnames]{xcolor}

\usepackage[colorlinks=true,linkcolor=blue,breaklinks=True,citecolor=brown,urlcolor=blue]{hyperref}

\usepackage{dingbat}

\usepackage{adjustbox}
\usepackage[linesnumbered,algo2e,ruled]{algorithm2e}
\usepackage{multirow}
\usepackage{booktabs,subcaption,dcolumn}
\usepackage{tikz}
\usepackage{enumitem}
\setlist[itemize]{noitemsep, topsep=0pt}
\usepackage{tabularx}
\usepackage[normalem]{ulem}

\usepackage{pifont}
\usepackage{svg}
\usepackage{soul}
\usepackage[font=footnotesize]{caption}
\usepackage[noadjust]{cite}

\usepackage{amsmath}
\usepackage{amsfonts, amssymb}
\usepackage{colortbl}
\usepackage{tablefootnote}
\usepackage[most]{tcolorbox}
\usepackage[sort&compress, numbers]{natbib}

\usepackage{siunitx}
\usepackage[table]{xcolor}
\usepackage{setspace}

\AtBeginDocument{%
  \providecommand\BibTeX{{%
    \normalfont B\kern-0.5em{\scshape i\kern-0.25em b}\kern-0.8em\TeX}}}

\newcolumntype{?}{!{\vrule width 1.5pt}}

\newcommand{\concap}[1]{{{\fontfamily{qcr}\selectfont C\MakeLowercase{on}C\MakeLowercase{ap}}}}
\newcommand{\command}[2]{{#2{\fontfamily{qhv}\selectfont #1}}}

\newcommand{\textbox}[1]{
    \noindent\fbox{%
        \parbox{0.97\columnwidth}{%
            {#1}
        }%
    }
}

\newtcolorbox{cooltextbox}[1][]{%
    colback=black!5,
    colframe=black!5,
    notitle,
    sharp corners,
    borderline west={0pt}{0pt}{red!80!black},
    enhanced,
    breakable,
    left=0pt,
    right=0pt,
    top=0pt,
    bottom=0pt
    }

\newcommand{\overbar}[1]{\mkern 1.5mu\overline{\mkern-1.5mu#1\mkern-1.5mu}\mkern 1.5mu}

\newcommand\smamath[1]{{\small $#1$}}
\newcommand\smacal[1]{{\small $\mathcal{#1}$}}

\newcommand\scmath[1]{{\scriptsize $#1$}}

\newcommand\giovanni[1]{%
  \bgroup
  \hskip0pt\color{red!80!black}%
  GIOVANNI: #1%
  \egroup
}
\newcommand\miel[1]{%
  \bgroup
  \hskip0pt\color{orange!80!black}%
  MIEL: #1%
  \egroup
}

\newcommand\revision[1]{\bgroup\hskip0pt\color{blue}#1\egroup}

\newcommand\revisiongreen[1]{%
 \bgroup
  \hskip0pt\color{black!50!green}%
 #1%
 \egroup
}

\begin{document}

\title{\concap{}: Practical Network Traffic Generation for (ML- and) Flow-based Intrusion Detection Systems}

\author{
\hspace{-0cm}
\IEEEauthorblockN{Miel Verkerken\IEEEauthorrefmark{2}, Laurens D'hooge\IEEEauthorrefmark{2}, Bruno Volckaert\IEEEauthorrefmark{2}, Filip De Turck\IEEEauthorrefmark{2}, Giovanni Apruzzese\IEEEauthorrefmark{5}
\\}
\IEEEauthorblockA{{ 
\IEEEauthorrefmark{2}\textit{Ghent University -- imec},
\IEEEauthorrefmark{5}\textit{University of Liechtenstein}, \IEEEauthorrefmark{5}\textit{Reykjavik University},
}}

{
\{name.surname\}@\{\IEEEauthorrefmark{2}ugent.be,
\IEEEauthorrefmark{5}uni.li\} 

}
}

\pagestyle{plain}
\maketitle

\begin{abstract}

\noindent
Network Intrusion Detection Systems (NIDS) have been studied in research for almost four decades. Yet, despite thousands of papers claiming scientific advances, a non-negligible number of recent works suggest that the findings of prior literature may be questionable. At the root of such a disagreement is the well-known challenge of obtaining data representative of a real-world network---and, hence, usable for security assessments. 

We tackle such a challenge in this paper. We propose \concap{}, a practical tool meant to facilitate experimental research on NIDS. Through \concap{}, a researcher can set up an isolated and lightweight network environment and configure it to produce network-related data, such as packets or NetFlows, that are automatically labeled---hence ready for fine-grained experiments. \concap{} is rooted on open-source software and is designed to foster experimental reproducibility across the scientific community by sharing just one configuration file. Through comprehensive experiments on 10 different network activities, further expanded via in-depth analyses of 21 variants of two specific activities and of 100 repetitions of four other ones, we empirically verify that \concap{} produces network data resembling that of a real-world network. We also carry out experiments on well-known benchmark datasets as well as on a real ``smart-home'' network, showing that, from a cyber-detection viewpoint, \concap{}'s automatically-labeled NetFlows are functionally equivalent to those collected in other environments. Finally, we show that \concap{} enables to safely reproduce sophisticated attack chains (e.g., to test/enhance existing NIDS). Altogether, \concap{} is a solution to the ``data problem'' that is plaguing NIDS research.

\end{abstract}

\begin{IEEEkeywords}
dataset, assessment, netflow, training, nids, cve, machine learning, replication, container, kubernetes, labeling
\end{IEEEkeywords}

\section{Introduction}
\label{sec:introduction}

\noindent
There is an anomaly in research on Network Intrusion Detection Systems (NIDS).
On the one hand, a deluge of papers continues to expand the boundaries of our knowledge, as shown by~\cite{khraisat2019survey, apruzzese2023sok, ahmad2021network}. On the other hand, a growing number of recent efforts highlight some ``pitfalls'' that undermine the foundations of prior research (e.g.,~\cite{arp2022and,engelen2021troubleshooting, liu2022error, catillo2023machine, flood2024bad}). 

To portray this contrast, we can look at the panorama of open-source datasets used in NIDS-related research, some of which count thousands of citations according to Google Scholar~\cite{flood2024bad}. As a concrete example, consider the paper presenting the CICIDS17 dataset~\cite{sharafaldin2018toward}: published in 2018, this paper had 1600 citations in Q3 2022, which increased to 4200 in Q4 2024---indicating a substantial growth in NIDS research. Yet, in 2021, Engelen et al.~\cite{engelen2021troubleshooting} pinpointed glaring issues in CICIDS17---particularly in terms of \textit{ground-truth labeling of attack samples}. The flaws of CICIDS17 (and also of its successor, CICIDS18) have been ``fixed'' in 2022~\cite{liu2022error}. However, the EuroS\&P'24 Best Paper Award~\cite{flood2024bad} revealed that most datasets used by prior research have ``bad design smells.'' 

Simply put, the stark reality is that 
{\small \textit{(i)}}~data is required to support a paper's claims, but {\small \textit{(ii)}}~high-quality data is hard to come by in the NIDS context---especially from the viewpoint of an academic researcher~\cite{apruzzese2023sok, flood2024bad, cordero2021generating}. To aggravate this problem, modern NIDS increasingly rely on data-driven techniques, such as machine learning (ML). Therefore, \textit{labeled} data is necessary to properly evaluate the pros and cons of state-of-the-art NIDS~\cite{apruzzese2023sok}. Finally, despite the benefits that open-source datasets (under the assumption that they are correctly labeled) can provide to a researcher~\cite{bonninghausen2024introducing}, exclusive reliance on such ``static benchmarks'' prevents one from \textit{generating new data}. This impairs the assessment of existing NIDS against recent, and more sophisticated threats---such as multi-step attacks~\cite{navarro2018systematic,mitreattack}, or attacks exploiting recently-discovered vulnerabilities~\cite{cve}. We aim to rectify such a ``data problem.''

Our major technical contribution is \concap{}, an open-source system to \textbf{generate (and label) network-traffic data mimicking that of a real-world network}. \concap{} is particularly suited to generate network data pertaining to \textit{malicious} activities. Such data can then be used alongside ``benign'' data taken from the network environment that the NIDS is meant to protect---which is a well-founded assumption~\cite{arnaldo2019holy, apruzzese2022cross, cordero2021generating}. Practically, such environment can be: {\small \textit{(a)}}~that of a benchmark dataset~\cite{bonninghausen2024introducing}, or of {\small \textit{(b)}}~an ad-hoc network testbed for experimental research~\cite{gomez2023survey}, or even that of {\small \textit{(c)}}~a real-world network~\cite{apruzzese2022role}. We design \concap{} so that it is {\small \textit{(i)}}~flexible enough to enable reproduction of any (allegedly malicious) network activity, whose generated datapoints are {\small \textit{(ii)}}~automatically labeled at the granular level, while also enabling
{\small \textit{(iii)}}~control of the network conditions (e.g., to simulate resource starvation). 
As such, \concap{} represents a solid foundation to address the ``data problem'' that affects NIDS research. {We will demonstrate this.}

{To develop \concap{}, we assembled open-source technologies, such as container-related frameworks, networking utilities, and network flows (NetFlow~\cite{apruzzese2023sok, verkerken2025rustiflow}) extractors. Our design \textit{facilitates experimental reproducibility}: to (automatically) generate labeled NetFlows pertaining to one (or more) attacks, \concap{} only requires the researcher to define a ``scenario,'' i.e., a file describing the activities of the ``attacker'' and ``target'' host(s), as well as the characteristics of the overarching network channel. In this way, other researchers can replicate the same experiments just by running the same ``scenario'' in their own version of \concap{}, thereby \textit{removing the need to share data}---which can be quite big in size (e.g., the CICIDS18~\cite{sharafaldin2018toward} dataset is larger than 400GB). Such a property intrinsically fosters scientific reproducibility---which is lacking even in top-tier security venues~\cite{olszewski2023get,apruzzese2023sok}.

We empirically validate the realism of \concap{}'s generated data. Through extensive experiments wherein we compare the network traffic produced by a physical bare-metal environment with that generated by \concap{}, we find that there are negligible differences (at both the packet- and NetFlow-level) between these two setups. We deeply investigate such differences, and confirm that they are due to the well-known ``unpredictable'' behavior of modern networks~\cite{sommer2010outside,apruzzese2023sok}. Inspired by such a finding, we go one step further and examine if the data generated by \concap{} is deterministic---an aspect that received limited attention from related research (e.g.,~\cite{clausen2019traffic}). Our analysis, entailing 100 executions of the same set of four distinct network activities, reveals that \concap{}'s output cannot be claimed to be fully deterministic---as is the case for real-world networks, too. However, through other experiments on well-known benchmarks and on a real ``smart-home'' network, we show that, ML-wise, the data generated by \concap{} is functionally equivalent to that of other network setups. 

Altogether, our evaluations demonstrate that \concap{} represents a convenient tool to address the ``data problem'' that affects NIDS-related research. For instance, we also show how to use \concap{} to carry out security assessments of NetFlow- and ML-based NIDS against recent CVE; as well as how to replicate a sophisticated attack chain, resembling nine MITRE ATT\&CK tactics, and generate the corresponding (labeled) traffic. Finally, we also show that \concap{} is lightweight: it takes less than 3 seconds to power-up on a laptop, and its memory footprint takes less than 40MB.

\textbf{\textsc{Contributions.}} After positioning our paper within extant literature and motivating our work (in Section~§\ref{sec:background}), we:
\begin{itemize}[leftmargin=*]
    \item Propose \concap{}, a system for generating ad-hoc network traffic~(§\ref{sec:concap}), particularly suited to \textit{create and label} malicious datapoints, facilitating network security research; we extensively describe and justify our implementation choices---rooted in open-source software and reproducibility (§\ref{sec:creation}).  
    \item Empirically demonstrate (§\ref{sec:validation}) that \concap{} generates network-traffic data resembling that of bare-metal setups, and prove that experiments run via \concap{} are not fully deterministic---a behavior expected by real-world networks.
    \item Show that (§\ref{sec:application}), from an ML viewpoint, the (labeled) NetFlows produced via \concap{} are functionally equivalent to those originating from other setups---such as from benchmark datasets, or from a real ``smart home'' network. We also show that \concap{} can generate data of sophisticated multi-step attacks and recent threats to test existing NIDS.
\end{itemize}
We also discuss our findings (§\ref{sec:discussion}) and compare \concap{} with related works (§\ref{ssec:related}). We release all our resources~\cite{repository}.

}
\section{Related Work and Motivation}
\label{sec:background}

\noindent
We summarize the domain of NIDS~(§\ref{ssec:nids}) and we outline the challenges of acquiring data for assessing NIDS (§\ref{ssec:obtaining}), representing the root of the problem tackled by our paper. Then, we motivate our technical (§\ref{ssec:generating}) contribution, which specifically focuses on NIDS analysing network flows (NetFlows~\cite{vormayr2020my}) due to the widespread usage of this datatype~\cite{apruzzese2023sok,flood2024bad,dias2020go}. 

\subsection{Network Intrusion Detection Systems (NIDS)}
\label{ssec:nids}

\begin{figure}[t]
\centering
\includegraphics[width=0.42\textwidth]{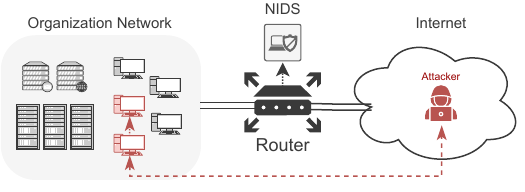}
\vspace{-0.1in}
\caption{
\textbf{Exemplary deployment scenario of an NIDS.}}
\label{fig:nids}
\vspace{-0.2in}
\end{figure}

\noindent
Modern networks are constantly under attack~\cite{enisa2023threat,cloudflare2024threat}, and NIDS represent the first line of defense against the ever-evolving cyberthreats~\cite{sans2024report}. The main goal of an NIDS\footnote{Borrowing from~\cite{apruzzese2023sok}, we use ``NIDS'' in a broad sense, including also SIEM~\cite{comodo2023idssiem} or EDR~\cite{paloalto2023edrsiem} (which can be seen as extensions of NIDS).} is to \textit{detect} any given threat (e.g., a botnet-infected host, a remote DDoS attack, or an attacker who acquired access to an internal host) as early as possible so that proper mitigations can be enacted to mitigate the damage of an offensive campaign~\cite{bertino2021ai}. Fig.~\ref{fig:nids} shows a schematic of an NIDS deployment scenario. 

The advent of data-driven technologies, such as AI/ML (which support both signature- and anomaly-based methods~\cite{apruzzese2023sok}), has been adopted by NIDS developers~\cite{kshetri2021economics}. Yet, even state-of-the-art NIDSs struggle with the sheer amount of attacks that target current organizations~\cite{enisa2023threat}. Intriguingly, even though practitioners do appreciate the analytical capabilities of ML~\cite{mink2023everybody}, modern security operation centers are overwhelmed with false alarms~\cite{alahmadi202299,vermeer2023alert}. Put simply, despite being an instrumental security tool, there is a constant need to improve NIDS---creating a fertile ground for research.

Since the seminal work by Denning~\cite{denning1987intrusion}, thousands of scientific papers have sought to propose (e.g.,~\cite{mirsky2018kitsune,barradas2021flowlens, piskozub2021malphase, fu2021realtime}), enhance (e.g.,~\cite{sommer2003enhancing,araujo2019improving}), or assess (e.g.,~\cite{apruzzese2022cross,jacobs2022ai,corona2013adversarial}) NIDS. Despite all such efforts, a recent SoK~\cite{apruzzese2023sok} revealed that there is skepticism among industry experts with regard to the results claimed in research. Such doubts are well-founded: various recent papers (e.g.,~\cite{liu2022error, catillo2023machine, flood2024bad}) identified critical issues that are becoming endemic in research. The root cause of most such critiques is the \textit{poor quality of the data} used to test these systems---which is a problem that has been known to affect this research domain since at least 2010~\cite{sommer2010outside}.

\subsection{Research Challenge: Obtaining Data for NIDS}
\label{ssec:obtaining}

\noindent
Any security tool must be tested before its deployment. Such tests require data. For NIDS, the evaluation should be carried out on data that is representative of the environment in which the NIDS is meant to be deployed~\cite{apruzzese2023sok}. 
Unfortunately, carrying out the abovementioned operations is tough from the perspective of an (academic) researcher. Let us outline the options that enable one to collect network-related data for a security assessment of a NIDS, explaining their challenges.
\begin{itemize}[leftmargin=*]
    \item \textit{Real-world capture from the deployment environment.} This is ideally the best option since it guarantees that the data resembles the one that will be generated by the monitored network. However, such an option may not be available: the researcher may not have access to such data due to privacy reasons, and infecting/attacking physical devices may not be acceptable in some organisations (even if for research~\cite{wangr2025rr}).

    \item \textit{Synthetic capture from a custom environment.} This is a sensible alternative: by creating an ad-hoc network (e.g., via virtual machines~\cite{sharafaldin2018toward}), a researcher has plenty of freedom to collect and generate any sort of data. However, the \textit{benign} data may not be representative of the deployment environment. Moreover, even in such a setup, it is currently challenging to precisely distinguish benign from malicious data points: as shown in~\cite{liu2022error, flood2024bad}, there is a risk of mistakenly ``label'' benign samples as malicious. Such errors can skew the results of the final assessment~\cite{krauss2024verify}. 
    
    \item \textit{Reliance on benchmark datasets.} The last option is to use publicly available data, e.g., generated by other researchers in their own environments (either from physical or virtualized setups). This is a convenient option: it is exempt from privacy concerns and requires minimal technical expertise (since no simulation occurs). However, the validity of the corresponding evaluation will depend on whether the benchmark is a meaningful representation of the network wherein the NIDS is meant to be deployed. 
\end{itemize}
For real-world deployments, it is paramount to evaluate the NIDS in the (real) network to be monitored by the NIDS: as highlighted by Sommer and Paxson~\cite{sommer2010outside}, networks present immense variability. Hence, even if any given NIDS is shown to ``work well'' on data from a custom network, it is questionable whether the same NIDS works well also in other networks. This is known as the ``generalisability''~\cite{verkerken2022towards,sarhan2022evaluating} (or ``transferability''~\cite{catillo2022transferability}) problem of NIDS, which prevents the creation of plug-and-play NIDS (confirmed by practitioners~\cite{apruzzese2022role}). However, \textit{a research paper needs not to aim for real-world deployment to provide a significant contribution to the state of the art}~\cite{apruzzese2023sok}. Our work is rooted in this truth: we focus on improving future research on NIDS---which~not~necessarily requires assessments on ``real'' networks to be valuable. 

\subsection{Practical Generation of Network Data}
\label{ssec:generating}

\noindent
Prior research on NIDS suffers from a ``data'' problem. Our main goal is to provide a solution to this problem by enabling future research to carry out meaningful assessments of NIDS. 

To elucidate the importance of our major contribution, we present a \textbf{motivational example}.
Suppose a researcher has no access to a real-world network for NIDS assessments. How can such a researcher conduct meaningful experiments? From our prior descriptions (§\ref{ssec:obtaining}) we identify three possible options.
\begin{itemize}[leftmargin=*]
    \item The researcher can create a simulated/virtual network, but doing so requires dealing with the labeling issues (for the malicious traffic), and is (likely) \textit{limited to small-scale evaluations} due to the impossibility of recreating a large network environment~\cite{clausen2019traffic}. 
    \item The researcher can exclusively rely on benchmark datasets, but this would \textit{limit the evaluation to the data within the benchmark dataset} (e.g., even by manipulating the benchmark's datapoints, there is a risk of breaking domain constraints~\cite{sheatsley2021robustness}), preventing exploration of new threats. 
    \item To overcome the abovementioned limitations, the researcher can do a mix of the above~\cite{arnaldo2019holy, apruzzese2022cross}: they can use ``benign'' data collected from a public dataset (potentially validated by prior work~\cite{liu2022error}), and then generate ``malicious'' data to use alongside the benign data to carry out a proper evaluation.
\end{itemize}
\textit{We argue that the third option is the most enticing one.} As of 2025, there are various publicly available datasets captured in large networks (see~\cite{bonninghausen2024introducing} for a list) whose data can be treated as ``benign'' for experimental purposes.\footnote{Such ``benign'' data can also come from the specific (real) network in which the NIDS is to be deployed---but, as we argued, this is not necessary (although it would increase the soundness of an evaluation).} Hence, if one could generate ``malicious'' data so that {\small \textit{(i)}}~it is ``correctly labeled,'' and {\small \textit{(ii)}}~it resembles the data generated by the host of the ``benign'' network, then one would be able to test a NIDS against a wide array of cyber threats---including new ones.

\vspace{1mm}
{\setstretch{0.99}
\textbox{ 
{\textsc{\textbf{Research Goal.}} We seek to develop an \textit{open-source} tool that enables the automatic \textit{generation} (and \textit{labeling)}} of network traffic data that resembles that of a real network.}
\begin{figure}[!t]
    \centering
    \includegraphics[width=\columnwidth]{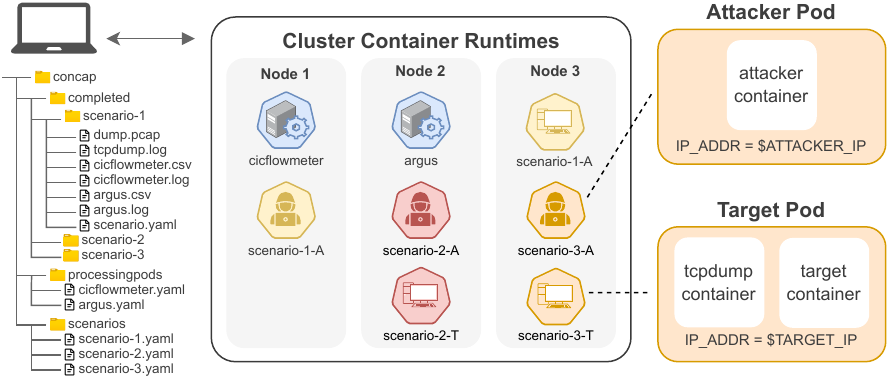}
    \vspace{-5mm}
    \caption{\textbf{Overview of \concap{}.} 
    \textmd{{\footnotesize [left] \concap{} configured with two NetFlow extractors and three scenarios, [mid] executing all scenarios simultaneously on the cluster. [right] A view of a running scenario's attacker and target pods.}}}
    \label{fig:concap-overview}
    \vspace{-5mm}
\end{figure}

\section{Our Proposed Tool: \textbf{\concap{\large}}}
\label{sec:concap}

\noindent
We present \concap{} (short for ``Container Capture''), our tool to generate realistic network traffic for research in NIDS. {After stating the design objectives~(§\ref{ssec:objectives}) and core principles~(§\ref{ssec:principles}) of \concap{}, we summarize its workflow~(§\ref{ssec:architecture}), schematically depicted in Fig.~\ref{fig:concap-overview}. We will discuss the low-level implementation of \concap{} in the next section (§\ref{sec:creation}}).

\subsection{Objectives}
\label{ssec:objectives}

\noindent
The underlying purpose of \concap{} is to simplify the operations involved in the generation and labeling of \textit{malicious} network activities that resemble those in a \textit{physical and real-world} network. \concap{} seeks to fulfill three objectives: 
\begin{itemize}[leftmargin=*]
    \item \textit{Attack flexibility.} The tool must allow one to specify and reproduce a variety of actions that lead to network communications---including those denoting ``malicious'' behavior. {The tool shall support both ``simple'' use cases (e.g., a machine attacking a different machine) as well as more ``sophisticated'' ones (e.g., a complex offensive campaign envisioning multi-step attacks across multiple hosts).}
    \item \textit{Data collection and labeling.} The tool must automatically: {\small \textit{(i)}}~capture network traffic related to the specified actions; {\small \textit{(ii)}}~generate statistical metadata summarizing the communications, i.e., NetFlows; {\small \textit{(iii)}}~assign a label to these NetFlows.
    \item \textit{Parallelism and control.} The tool must enable simultaneous execution of diverse experiments, and fine-grained control of the network conditions of each experiment---so to reproduce the behavior of any network, and enable ``bulk'' experiments.
\end{itemize}
{To align these objectives with our overarching research goal (see §\ref{ssec:generating}), the entirety of \concap{}'s source code (which must leverage open-source libraries/frameworks) is publicly released~\cite{repository}; moreover, the authors will make an effort to ensure long-term maintenance of \concap{}'s code repository.}

{
\subsection{Core Principles}
\label{ssec:principles}

\noindent
We outline the core principles of \concap{}, showing that our three objectives are embedded in its underlying design.}

\concap{} is rooted on the concept of ``scenario'', which serves as a customizable blueprint for an entire experiment. When defining a scenario, developers can specify: {\small \textit{(i)}}~the activities carried out by the involved hosts; 
{\small \textit{(ii)}}~the conditions of the overarching network environment;
{\small \textit{(iii)}}~the fine-grained label to assign to the generated network traffic metadata. 

\concap{} executes each scenario in an isolated ``containerized'' environment that mimics the behavior of real, physical hosts, and allows for unlimited parallel execution of scenarios---subject to the resource constraints of the experimental testbed on which \concap{} is executed. 

A scenario in \concap{} assumes a set of hosts: one designated as the ``attacker'', and one {or more} designated as the ``target''. {These hosts are fully controllable by the developer through customizable container configurations and attack commands. \concap{} captures the network packets (as a PCAP trace) exchanged between the attacker and the target(s), automatically extracts high-level NetFlow records, and then labels such NetFlows according to the scenario definition. Labeling accuracy is guaranteed by configuring the attacker host to only reproduce ``malicious'' actions, since the isolated environment ensures no background noise is present.

\concap{} supports reproducing both simple attacks, as well as complex, multi-step attack chains. Such support is provided via the ```scenario'' configuration file.
In simple-attack scenarios, the attacker host executes a sequence of actions against a single host; for instance, the scenario in Listing~\ref{lst:scenario-file} (in the Appendix~\ref{app:concap-configuration}) shows a simple port-scan done by one ``attacker'' host against one ``target'' host. In complex-attack scenarios, however, it is possible to designate multiple ``target'' hosts---which can be either used to reproduce an attack against multiple hosts (e.g., an horizontal port scan; or a multi-chain attack in which each step requires the completion of the previous steps). The network traffic of each target is captured individually and labeled according to the specifications provided in the scenario file, enabling fine-grained analysis.
}

\subsection{Architecture and Workflow}
\label{ssec:architecture}

\noindent
We present in Fig.~\ref{fig:concap-overview} a schematic of a typical setup of \concap{}, highlighting the most relevant logical units. We use Fig.~\ref{fig:concap-overview} to explain the operational workflow and architecture of \concap{}.

\textbf{Input.} On the left, the ``scenarios'' folder contains three YAML configuration files, each defining a given scenario to be executed by \concap{}. An example scenario configuration is shown in Listing~\ref{lst:scenario-file}, highlighting the various options (i.e., attacker and target definition, network conditions, and labeling) that can be configured before executing any given experiment. 
For added flexibility, \concap{} supports using different NetFlow generation tools. This can be specified through configuration files located in the “processingpods” folder, where the details of the desired tool are defined.

\textbf{Execution.}
When executing a scenario, \concap{} parses the configuration files and interacts with both the host machine and the cluster, consisting of one or more nodes. If multiple scenarios are specified, \concap{} supports concurrent execution by distributing them across different nodes, enabling parallelism (see the central section of Fig.~\ref{fig:concap-overview}).
\concap{} follows the instructions in the scenario files to set up the attacker and target pods, apply custom network configurations, initiate traffic captures, trigger attack execution, and run processing pods for feature extraction and labeling. See the rightmost section of Fig.~\ref{fig:concap-overview} for a detailed view of the attacker and a target pod during scenario execution.

\textbf{Output.}
For each scenario, \concap{} creates a dedicated folder containing all experiment outputs, including logs, the PCAP trace, and the labeled NetFlows (see left of Fig.~\ref{fig:concap-overview}). The reason why we focus (also) on NetFlows for \concap{} is due to their widespread popularity for network-related experiments (both in research and in practice~\cite{apruzzese2023sok,oh2022deepcoffea}); for instance, most public benchmark datasets are also released in this format~\cite{bonninghausen2024introducing}. Such a design choice serves to facilitate future research.

\section{Creation and Functionalities of \textbf{\concap{\large}}}
\label{sec:creation}

\noindent
We first explain and justify the key design choices followed to create \concap{}~(§\ref{ssec:design}). Then, we describe the ``scenario file,'' which embeds the most original contributions of \concap{}~(§\ref{ssec:scenario}). Finally, we explain how \concap{} enables also the simulation of sophisticated malicious activities, such as multi-step and multi-target attacks~(§\ref{ssec:advanced}).

\subsection{Design and Implementation Choices}
\label{ssec:design}

\noindent
{We present and justify the elements that compose \concap{}. 

\textbf{Isolated \textit{Containerized} Environment.}
As explained~(in §\ref{ssec:obtaining}), the major advantage of running network simulations is the ability to generate malicious network traffic without putting real-world networks at risk. Such simulations can leverage, e.g., virtual machines (as done, e.g., in~\cite{sharafaldin2018toward}) or \textit{containers}. 
Containers are a lightweight alternative to virtual machines~\cite{potdar2020performance}, which are becoming very popular in related research (e.g.,~\cite{clausen2019traffic, catillo2024towards, boettiger2015introduction}) also for the ease of reproducibility~\cite{moreau2023containers} {and reduced resource requirements~\cite{potdar2020performance}}.
For this reason, we used containers to develop \concap{}. 
There exist many open-source solutions that can be used to deploy containerized applications~\cite{casalicchio2020state}. For \concap{}, we rely on Kubernetes~\cite{kubernetes}, due to its widespread adoption ($>$110k GitHub stars~\cite{kubernetesGithub}) and key advantages over alternatives, such as enabling deployment and management of workflows across multiple machines~\cite{bentaleb2022containerization}---which is required to achieve our design goals of simultaneous execution of scenarios. For comparison, DetGen~\cite{clausen2019traffic} relies on Docker~\cite{docker}, which by default does not support parallel scenario execution across multiple machines {(we will compare \concap{} with DetGen in §\ref{ssec:comparisons})}. We stress, however, that Kubernetes alone \textit{does not} provide the functionalities provided by \concap{} (see §\ref{ssec:architecture}): we simply use Kubernetes as the backbone.

\textbf{Attacker and Target(s).}
In Kubernetes, a ``pod'' is the smallest deployable unit, functioning as a logical host that groups one or more containers with shared storage and networking resources. 
To ensure isolation and enable fine-grained control over each host involved in the ``attack'', we create \concap{} so that the attacker and target(s) hosts are deployed in separate pods. This allows to configure parameters such as bandwidth and latency for each host.
The attacker pod runs one container that simulates the behavior of the attacker's host. In contrast, each target pod runs two containers: one serving as the host targeted by the attacker, and another that captures the attacker-target network communications (via \command{tcpdump}{\small}{, which can capture all traffic of the target host since containers within the same pod share networking resources). 

{\textbf{NetFlow format.}} \concap{} supports any type of NetFlow generation software, {including novel ones (e.g.,~\cite{verkerken2025rustiflow}).} {Choosing a given NetFlow tool is done by editing a dedicated configuration file (shown in Listing~\ref{lst:processing-file} in the Appendix~\ref{app:concap-configuration}).
In our proposed implementation of \concap{}, we have integrated CICFlowMeter and Argus, which are popular in research and open source~\cite{vormayr2020my,apruzzese2023sok}.} {Recent surveys~\cite{apruzzese2023sok,verkerken2025rustiflow} revealed that most papers 
on ML-NIDS recently published in various (including top-tier) security venues rely on NetFlow for their analyses, motivating our focus on NetFlow. 

{
\textbf{Reproducibility.}
An implicit objective of \concap{}, from a scientific viewpoint, is to facilitate the reproducibility of network traffic generation experiments. Many existing public network traffic datasets lack detailed metadata/instructions about how the traffic was produced, often providing only high-level descriptions of the experimental setup~\cite{flood2024bad}. This lack of transparency makes it difficult to perform an in-depth analysis of such datasets and prevents researchers from linking individual network traffic to their specific causes or originators, which is critical for tasks such as automated detailed labeling. Reproducibility in this context means not only re-running the same scenario and obtaining consistent traffic captures and labels, but also \textit{sharing complete experimental configurations in a way that others can reliably replicate the results on different systems}.
To address this, \concap{} encapsulates all aspects of the traffic generation experiment in a reusable scenario file. These files define the behavior of the attacker and target(s), their environment, and the network conditions, serving as a blueprint of the experiment. By versioning and sharing these files alongside the resulting datasets, \concap{} enables other researchers to reproduce the traffic generation process with minimal manual setup and to understand the context behind the captured network traffic.
Moreover, the framework enforces consistency by automating the full execution pipeline--from deployment and traffic capture to flow extraction and labeling--ensuring that experiments can be rerun with similar results.}
}

\subsection{Customisability (Scenario definition)}
\label{ssec:scenario}

\noindent
Among \concap{}'s greatest advantages is its customisability---which we claim as our major original contribution. Indeed, we are not aware of any tool which provides the same degree of flexibility (we discuss related work in §\ref{ssec:related}). Such advantages are enabled by our custom implementation of the \textit{scenario} file, for which we have provided a snippet in Listing~\ref{lst:scenario-file}. In what follows, we provide more details on the functionalities provided by the scenario file. Recall that the scenario file follows a YAML notation which requires to specify four components: the attacker, the target, the network, and the label.}

{\textbf{Host configuration.}} The components describing the ``attacker'' and ``target'' hosts require a \textit{name} and the \textit{containerImage} to deploy the containers that simulate their behavior. Additionally, we designed \concap{} so that it is possible to specify the conditions of the computing runtime (e.g., \textit{CPU} and \textit{memory}): this is crucial for attacks that disrupt the availability of the target host (e.g., DoS). The ``attacker'' component also has the \textit{atkCommand} (used to start the attack), and an optional \textit{atkTime} parameter that controls the attack duration (in seconds): intuitively if \textit{$atkTime>0$}, then the attack will stop after the provided number of seconds; otherwise, the attack will continue until it naturally terminates. 
We also allow to configure the startup probe of the ``target'' component, which determines when the target is ready, as well as the filter used by \command{tcpdump}{\small} for the packet capture (PCAP). {Additionally, the attacker and/or target(s) can be run in privileged mode. This is useful for tools or services that need elevated permissions (e.g., raw socket access) or to simulate sophisticated attacks.}

{\textbf{Networking.} The \textit{network} component defines} the networking environment, enabling to specify, e.g., the bandwidth, latency, and packet loss of the communication channel. {Such a functionality is useful to, e.g., simulate various conditions---not only for covering a wide array of ``benign'' network environments, but also to reproduce circumstances which can conceal traces of malicious activities (e.g.,~\cite{catillo2024towards}). We are not aware of any open-source traffic-generation tool that enables (by default) specific configurations of the network conditions at the host level. This component, which we implemented via the open-source Traffic Control~\cite{tc}, also allows to determine if the generated traffic follows a specific distribution.}

{\textbf{Labeling.}} The \textit{label} component describes the labeling logic applied to the NetFlows (generated after processing the PCAP file). This can be configured globally, and individually per host. Given that \concap{} enables precise control of the entire attack workflow (i.e., attacker and target(s) host and network conditions), the assigned labels ensure the resulting NetFlows are associated with the correct ground truth (since they all share the same ``malicious'' generative process). {Moreover, since certain cyberattacks can be part of more sophisticated offensive campaigns (e.g., a port-scan can be part of a lateral-movement operation~\cite{apruzzese2017detection}), and since such use-cases are supported in \concap{}, it is possible to assign more granular labels. For instance, in Listing~\ref{lst:scenario-file} the ``category'' is ``port-scan'' whereas the subcategory is ``nmap'': to better identify the corresponding malicious traffic in a lateral-movement context, it would have been possible to specify ``lateral-movement'' as ``category'', and ``port-scan'' as ``subcategory''.}}

{
\subsection{Sophisticated attacks (multi-step and multi-host)}
\label{ssec:advanced}

\noindent
In its most simple form, \concap{} can be used to generate network traffic entailing one attacker and one target host. However, we designed \concap{} so to also enable simulation of more complex network activities, such as multi-host attacks (e.g., an attacker host seeking to scan multiple target hosts) or multi-step attacks (e.g., an attacker that seeks to brute force an SSH server after finding out that there is an active SSH service running on such a host) in a reproducible and structured way. Here, we describe how we enabled support of multi-step/host attacks in \concap{} (which we also claim as an original technical contribution).

\textbf{Challenges.} Simulating such advanced scenarios is non-trivial. Multi-step attacks often involve transferring state between phases (e.g., discovered credentials~\cite{apruzzese2017detection}) combined with dynamic network reconfiguration (e.g., proxy routing or port forwarding~\cite{angelini2019mad}). Additionally, these attacks may require accessing internal services only reachable after a successful compromise of a given gateway, demanding precise orchestration of timing, sequencing, and conditional logic. 

\textbf{Solutions.} To address these challenges, \concap{} introduces several mechanisms. 
Scenario definitions support dynamic environment variable substitution---instead of being constrained by hard-coded variables.
Moreover, \concap{}'s built-in startup-readiness probes ensure that attacks are launched only when the targets are fully operational.
Scenario-wide and per-target labels are automatically merged to annotate captured flows based on their context within the attack chain. 
Fig.~\ref{fig:concap-multi-target} illustrates the fully-automated process enabled by \concap{}. On the left, the developer defines an arbitrary and flexible attack chain using the ``atkCommand'' in the scenario file that specifies the attacker's behavior. In the middle, \concap{} automatically deploys the pods, applies the network configuration, and manages pod lifecycles and readiness. On the right, it performs per-target traffic capture, flow extraction, and labeling. This workflow is repeated for each custom-defined step of the attack chain. In a sense, \concap{} acts as a supervisor guaranteeing that each step is executed at the right time, thereby enabling smooth simulation of multi-step attacks---potentially requiring information acquired after intermediate steps. An end-to-end demonstration of using \concap{} to reproduce a multi-step/multi-target attack chain, resembling acknowledged MITRE ATT\&CK tactics, is provided in the Appendix~\ref{sapp:multistep}. 
}

\begin{figure}[!t]
    \centering
    \includegraphics[width=\columnwidth]{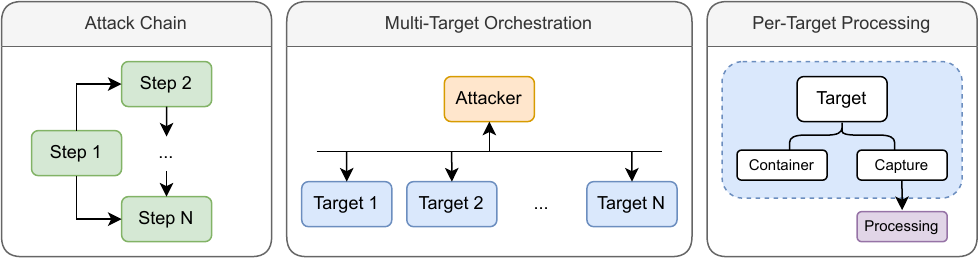}
    \caption{{\textbf{Multi-target scenario with \concap{}.}
    \textmd{{\footnotesize [left] \concap{} supports advanced multi-step attack chains, [mid] performed over multiple targets. [right] The traffic is captured and processed per target, enabling per-target labeling.}}}}
    \label{fig:concap-multi-target}
    \vspace{-4mm}
\end{figure}

\begin{cooltextbox}
\textsc{\textbf{Takeaway.}} \concap{} is the first open-source tool for generating realistic, labeled network traffic for NIDS research. It offers attack flexibility, automated NetFlow generation and labeling, and parallel scenario execution with fine-grained control. A short demo is available in our repository~\cite{repository}.
\end{cooltextbox}
\section{Real-world Validation of \concap{\large}}
\label{sec:validation}

\noindent
{The design choices we followed to develop \concap{} (e.g., the usage of containers, or the usage of Traffic Control) indicate that \concap{} is likely to generate traffic that mimics that of a physical real-world network. In what follows, we empirically confirm this hypothesis---a validation that has not been done in most prior works proposing traffic-generation tools (e.g.,~\cite{zhoutrafficformer}). Specifically, we ask ourself our first research question (RQ1):}

\vspace{1mm}
{\setstretch{0.99}
\textbox{{\textsc{\textbf{Research Question \#1:}} Does \concap{} generate network traffic that resembles that of a physical network?}
}

{\textbf{Motivation.} For instance, while implementing \concap{}, we may have made some mistakes; or it is possible that the usage of containers may lead to different traffic (w.r.t. that of a physical network). It could also be that there are some intrinsic differences: in which case, it is instructive to understand their root cause. Indeed, we are not aware of prior works on containers empirically testing such an hypothesis. 
Hence, our assessment serves to gauge whether our implementation of \concap{} is correct, and that \concap{} meets our overarching research goal (see §\ref{ssec:generating}); as well as studying potentially unknown properties of containers from a network perspective. 

\textbf{Approach.} To answer RQ1, we carry out an intuitive set of experiments: we compare the network data generated by \concap{} to that of a physical ``bare-metal'' setup---across a wide range of network activities. Such comparisons entail both quantitative and qualitative assessments, both at the network-packet level, as well as at the network-flow level. In particular, to comprehensively address RQ1, we perform three different experiments. First, as a starting point, we execute ten network activities, and see if there are any substantial differences from a quantitative viewpoint~(§\ref{ssec:broad}). Second, we conduct an in-depth analysis focused on examining over twenty different variants of two specific activities types (§\ref{ssec:indepth}). Third, as an ancillary experiment, we study the extent to which the traffic generated by \concap{} can be considered to be deterministic~(§\ref{ssec:deterministic}).

\textbf{Common Setup and Workflow.} The experiments discussed in this section entail two distinct environments.} 
\begin{itemize}[leftmargin=*]
    \item \textit{Physical network.} Two ``bare-metal'' physical hosts, each having six Intel Core i5-9400, 32GB RAM running Ubuntu 20.04.6 and connected by a 1 Gbit switch.
    \item {\concap{}. A Kubernetes cluster (v1.29.0) with 1 control plane and 3 worker nodes. Each node has 16GB RAM, four Intel Xeon E5-2640v4, running Ubuntu 22.04.3. The machines are interconnected by a 10 Gbit switch.}
\end{itemize}
{For each experiment, we follow a similar procedure. We first consider the physical network. We instruct one bare-metal host, acting as the ``attacker'', to carry out each of the attacks envisioned in the experiment; whereas the other host acts as the ``target'' and is configured to enable the successful execution of the attack (e.g., if the attack is an SSH bruteforcing, the ``target'' will be running an SSH server). 
The corresponding traffic is captured (as PCAP) on the ``target'' host via \command{tcpdump}{\small}, and we also generate the corresponding network flows (via CICFlowMeter and Argus).
Then, we consider the \concap{} environment: we configure \concap{} so that the ``attacker'' and ``target'' hosts resemble (in terms of computational power, software, and commands executed) of the ``bare metal'' machines; this ensures that the only ``variable'' across our experiments is the overarching environment (i.e., \concap{} or the physical setup). The configuration files are in our repository~\cite{repository}.

\begin{table}[!t]
    \caption{{\textbf{Broad analysis of various network activities.} \textmd{{\footnotesize We qualitatively compare the number of packets and NetFlows generated by our bare-metal and \concap{} environments.  Network traffic on bare-metal servers and \concap{}.}}}}
    \vspace{-1mm}
    \label{tab:broad}
    \centering
    \resizebox{\columnwidth}{!}{
        {
        \begin{tabular}{l S[table-format=6] S[table-format=6] S[table-format=6] S[table-format=6] S[table-format=6] S[table-format=6]}
        \toprule
        \multirow{2}{*}{\parbox{1.5cm}{\textbf{Tool}}} & \multicolumn{2}{c}{\textbf{Number of Packets}} & \multicolumn{2}{c}{\textbf{CICFlowMeter Flows}} & \multicolumn{2}{c}{\textbf{Argus Flows}} \\
        \cmidrule{2-7}
        & \textbf{Bare-metal} & \textbf{ConCap} & \textbf{Bare-metal} & \textbf{ConCap} & \textbf{Bare-metal} & \textbf{ConCap} \\
        \midrule
        \command{ping}{\small} & 20 & 20 & 1 & 1 & 10 & 10 \\
        \command{dig}{\small} & 10 & 10 & 5 & 5 & 5 & 5 \\
        \command{mysql}{\small} & 37 & 27 & 1 & 1 & 1 & 1 \\
        \command{curl/ftp}{\small} & 4910 & 1675 & 2 & 2 & 2 & 2 \\
        \command{nmap}{\small} & 5 & 5 & 2 & 2 & 2 & 2 \\
        \command{nmap (-sV)}{\small} & 128 & 103 & 10 & 10 & 11 & 11 \\
        \command{patator (SSH)}{\small}  & 30960 & 31485 & 680 & 680 & 680 & 680 \\
        \command{patator (FTP)}{\small} & 2093 & 1860 & 70 & 70 & 70 & 70 \\
        \command{slowloris}{\small} & 21300 & 21305 & 1500 & 1500 & 1500 & 1500 \\
        \command{wfuzz}{\small} & 2310 & 2086 & 15 & 15 & 15 & 15 \\
        \bottomrule
        \end{tabular}
    }}
    \vspace{-4mm}
\end{table}

\subsection{Broad Analysis (many different network activities)}
\label{ssec:broad}

\noindent
We begin our quest to answer RQ1 with a broad, preliminary analysis of quantitative nature. Specifically, we use our environments to reproduce ten network activities of various type, capture the corresponding traffic and generate the NetFlows, and quantitatively compare the overall number of packets and NetFlows generated by both environments.

\textbf{Network Activities.} For a comprehensive assessment, we consider activities of both ``benign'' and ``malicious'' nature. Four are clearly malicious: 
\command{wfuzz}{\small} (an exemplary HTTP fuzzying attack~\cite{wfuzz});
\command{slowloris}{\small} (an exemplary DoS attack~\cite{slowloris});
as well as two variants of \command{patator}\small{} (an exemplary bruteforcing attack), one focused on SSH- and another on FTP-bruteforcing. Whereas the other activities are common network-related commands, not necessarily of a malicious nature: 
\command{ping}{\small}~\cite{ping}, \command{dig}{\small}~\cite{dig}, \command{mysql}{\small}~\cite{mysql}, \command{curl}{\small}~\cite{curl} over \command{ftp}{\small}~\cite{ftp}, as well as two variants of \command{nmap}{\small}~\cite{nmap}. Such a broad set of activities is hence a solid basis to investigate RQ1 at a high level.

\textbf{Results.} We execute each activity with ``default'' options (we report the exact commands and configurations in the Appendix~\ref{sapp:commands-concap}). For each activity, we capture the corresponding packets and generate the respective NetFlows. We report in Table~\ref{tab:broad} the results of this experiment. Specifically, we report the total number of packets, as well as the total number of NetFlows (generated both via CICFlowMeter and Argus) of each activity for the physical and \concap{} environments. We can already see that the number of NetFlows is a perfect match, which is a promising result in the context of answering RQ1 positively. Regarding the number of packets, we see that only three activities (\command{ping}{\small}, \command{nmap}{\small}, \command{dig}{\small}) had an equal number of packets. In the next experiment, we will better examine two activities for which we found differences: the SSH variant of \command{patator}{\small}, and \command{nmap} in its service/version probe variant.

}

{
\subsection{In-depth Analysis (multiple variants of two activities)}
\label{ssec:indepth}

\noindent
Our previous experiment (§\ref{ssec:broad}) revealed that even though \concap{} produced nearly-identical traffic (quantitatively-wise) to that of a physical network environment, some network activities (e.g., \command{ssh-patator}{\small} and \command{nmap}{\small}) presented some differences. Here, we better examine these phenomena.

\subsubsection{\textbf{Methodology}}
\label{sssec:method}
For a deep understanding, we expand our previous assessment and consider 21 variants of the aforementioned activities---up from 2 (i.e., the default configurations). Specifically, for \command{nmap}{\small}, we consider the 12 possible combinations of the scanning options: \command{-Pn -sS -sV -sT -sU}{\small}. Whereas for ssh-\command{patator}{\small} we test 9 combinations by varying: \command{persistent=0/1 -RL=0/1 -T=1/5/10}{\small}. Note that, for both \command{patator}{\small} and \command{nmap}{\small}, our variants also include the default option---which we repeat for completeness.
After capturing the PCAP trace and generating the NetFlows, we first examine differences (between \concap{} and the physical environment) at the packet level: to this end, we use Wireshark~\cite{wireshark} for qualitative assessments, and then quantitatively inspect the number of packets generated by each variant of our considered activities. Finally, we quantitatively examine differences at the NetFlow level.

\subsubsection{\textbf{Packet-level analysis}}
\label{sssec:packetlevel}
After qualitatively inspecting the PCAP traces, we found that, across the 12 \command{nmap}{\small} and the 9 \command{patator}{\small} variants, the bytes in the individual packets generated by \concap{} \textit{exactly match those of the physical network}---except for expected differences in headers (e.g., MAC and IP addresses, high ports, checksums). We also observed some variations in the \textit{window size} and \textit{maximum segment size} which affect the amount of data that can be handled by the receiver: such (minimal) differences are due to the physiological diversity of each network, and their existence is evidence that the packets generated by \concap{} can also present a degree of uniqueness which is intrinsic to physical real-world networks.
}

\begin{table}[!t]
    \caption{\textbf{Network traffic on bare-metal servers and \concap{}.} \textmd{\footnotesize 
    For both network activities, the number of NetFlows is identical and the number of network packets has minimal variations (as expected in realistic networks).}}
    \label{tab:realistic-traffic}
    \begin{subtable}{\columnwidth}
        \centering
        \resizebox{\columnwidth}{!}{
            \begin{tabular}{lrrrrrr}
            \toprule
            \multirow{2}{*}{\parbox{1.5cm}{\textbf{Attack Options}}} & \multicolumn{2}{c}{\textbf{Number of Packets}} & \multicolumn{2}{c}{\textbf{CICFlowMeter Flows}} & \multicolumn{2}{c}{\textbf{Argus Flows}} \\
            \cmidrule{2-7}
            & \textbf{Bare-metal} & \textbf{ConCap} & \textbf{Bare-metal} & \textbf{ConCap} & \textbf{Bare-metal} & \textbf{ConCap} \\
            \midrule
            -Pn -sS & 5 & 5 & 2 & 2 & 2 & 2 \\
            -Pn -sS -sV & 122 & 103 & 10 & 10 & 11 & 11 \\
            -Pn -sT & 6 & 6 & 2 & 2 & 2 & 2 \\
            -Pn -sT -sV & 127 & 107 & 10 & 10 & 11 & 11 \\
            -Pn -sU & 4 & 4 & 3 & 3 & 4 & 4 \\
            -Pn -sU -sV & 4 & 4 & 3 & 3 & 4 & 4 \\ 
            -sS & 13 & 13 & 5 & 5 & 6 & 6 \\
            -sS -sV & 137 & 113 & 13 & 13 & 15 & 15 \\
            -sT & 14 & 14 & 5 & 5 & 6 & 6 \\
            -sT -sV & 133 & 115 & 13 & 13 & 15 & 15 \\
            -sU & 12 & 12 & 6 & 6 & 8 & 8 \\
            -sU -sV & 12 & 12 & 6 & 6 & 8 & 8 \\ 
            \bottomrule
            \multicolumn{7}{l}{-Pn = Treat host as online} \\
            \multicolumn{7}{l}{-sS / -sT / -sU = TCP SYN, TCP Connect or UDP scan} \\
            \multicolumn{7}{l}{-sV = Probe open ports to determine service and version info} \\
            \end{tabular}
        }
       \caption{Nmap port scan}
       \label{tab:realistic-nmap}
    \end{subtable}
    \hfill
    \begin{subtable}{\columnwidth}
        \centering
        \resizebox{\columnwidth}{!}{
            \begin{tabular}{l S[table-format=6] S[table-format=6] S[table-format=4] S[table-format=4] S[table-format=4] S[table-format=4]}
            \toprule
            \multirow{2}{*}{\parbox{1.5cm}{\textbf{Attack Options}}} & \multicolumn{2}{c}{\textbf{Number of Packets}} & \multicolumn{2}{c}{\textbf{CICFlowMeter Flows}} & \multicolumn{2}{c}{\textbf{Argus Flows}} \\
            \cmidrule{2-7}
            & \textbf{Bare-metal} & \textbf{ConCap} & \textbf{Bare-metal} & \textbf{ConCap} & \textbf{Bare-metal} & \textbf{ConCap} \\
            \midrule
            P=0 RL=0 T=1 & 95194 & 78309 & 3400 & 3400 & 3400 & 3400 \\
            P=1 RL=0 T=1 & 33103 & 29698 & 578 & 578 & 578 & 578 \\
            P=1 RL=1 T=1 & 38726 & 35332 & 578 & 578 & 578 & 578 \\
            P=0 RL=0 T=5 & 95170 & 78426 & 3400 & 3400 & 3400 & 3400 \\
            P=1 RL=0 T=5 & 33554 & 30293 & 595 & 595 & 595 & 595 \\
            P=1 RL=1 T=5 & 39066 & 35823 & 595 & 595 & 595 & 595 \\
            P=0 RL=0 T=10 & 94941 & 78347 & 3400 & 3400 & 3400 & 3400 \\ 
            P=1 RL=0 T=10 & 35399 & 31506 & 680 & 680 & 680 & 680 \\
            P=1 RL=1 T=10 & 40772 & 37384 & 680 & 680 & 680 & 680 \\
            \bottomrule
            \multicolumn{7}{l}{ P = Persistent, RL = Rate-Limit, T = Threads} \\
            \end{tabular}
        }
        \caption{Patator SSH Bruteforce}
        \label{tab:realistic-patator}
     \end{subtable}
     \vspace{-6mm}
\end{table}

{Next, we focus on the \textit{total number of packets} exchanged for each activity. We report the results in Tables~\ref{tab:realistic-nmap} (for \command{nmap}{\small}) and~\ref{tab:realistic-patator} (for \command{patator}{\small}). The leftmost column reports the specific options for each attack, whereas the second and third columns report the packets exchanged by the ``target'' and ``attacker'' host during the attack for both the physical and \concap{} environment (we also report the NetFlows, covered in §\ref{sssec:netflowlevel}). 
\begin{itemize}[leftmargin=*]
    \item \command{nmap}{\small}. For the port scan, there is an almost perfect match. The differences occur only when the attacker probes the service running on the open port (option \command{-sV}{\small}). This is expected: the \command{-sV}{\small} option induces the server to provide the index HTML page of the Apache webserver, which is 10,918 bytes. In \concap{}, such an exchange requires 2 packets, whereas the same payload requires 8 packets in the physical network---due to small differences in the network conditions such as TCP window scale~\cite{rfc813, rfc7323}.

    \item \command{patator}{\small}. For the SSH brute force, we observe differences of \smamath{\approx}\smamath{10\%} in the number of packets. In this case, the difference is due to the TCP/IP stack handling data acknowledgment on the different hardware setups. Additional testing in another replication experiment on a second real network showed similar but different deviations in the number of ACKs. 
    The TPC RFC~\cite{rfc793} allows for this nondeterministic behavior of ``delayed ACKs'' which send fewer than one ACK segment per data segment received and is even expected by the official specification~\cite{rfc813, rfc1122} to increase efficiency in the Internet and the hosts. However, as demonstrated by our qualitative analysis, the payload is the same.
   
\end{itemize}
In summary, any packet-level differences are due to inherent and unpredictable characteristics of the network, which do not affect the contents of the communication payloads.

\subsubsection{\textbf{NetFlow-level analysis}}
\label{sssec:netflowlevel}
First, we carry out a quantitative comparison focusing on the number of NetFlows exchanged between the ``attacker'' and ``target'' host for each activity---reported in Tables~\ref{tab:realistic-nmap} and~\ref{tab:realistic-patator}. Notably, there is always a perfect match: despite differences in packet counts, the number of NetFlows is consistent when the PCAP is processed by the same NetFlow software (CICFlowMeter and Argus follow a different logic to create NetFlows\footnote{\textbf{Bugfix:} we encountered an unexpected discrepancy in the NetFlows generated by Argus. We reached out to the developers, and they confirmed that there was a bug in \textit{their} implementation, and we helped fixed their code. After fixing the code, the number of NetFlows is identical.}). 
Then, we analyze the distributions of NetFlow features (for simplicity, we focus only on CICFlowMeter) across the different attacks. Fig.~\ref{fig:feature-dist-bare-concap} presents side-by-side comparisons of the \textit{mean packet length} distribution for all variations of our attacks. We also report the plots for \textit{all 30 NetFlow features} in Fig.~\ref{fig:feature_dist} (in the Appendix~\ref{app:results}). We observe that all features exhibit similar distributions, or any differences can be explained via our packet-level analyses (e.g., more packets sent with empty payload lead to a decrease in the mean packet length). 
}

\begin{figure}[!t]
    \centering
    \begin{subfigure}[b]{0.48\linewidth}
         \centering
        \includegraphics[width=\linewidth]{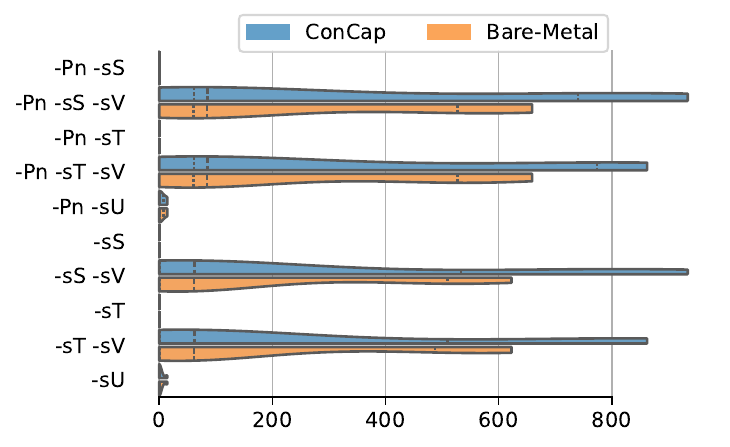}
    \end{subfigure}
    \hfill
    \begin{subfigure}[b]{0.48\linewidth}
         \centering
        \includegraphics[width=\linewidth]{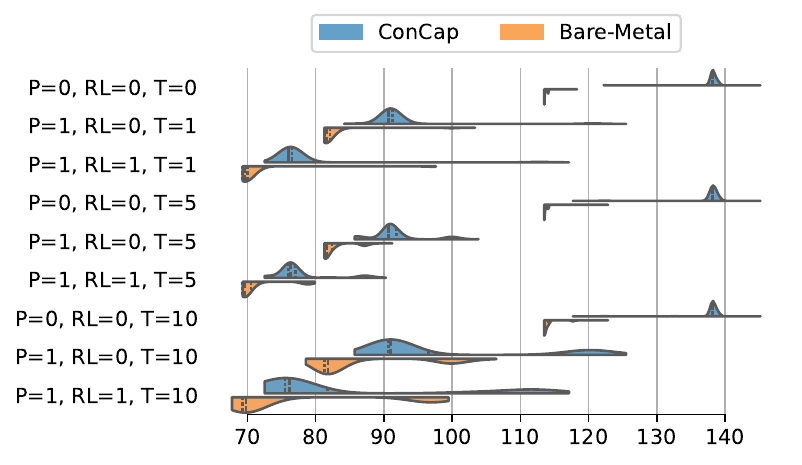}
    \end{subfigure}
    \vspace{-1mm}
    \caption{\textbf{NetFlow feature distribution of \textit{mean packet length} for traffic generated by \concap{\scriptsize} and a pair of physical hosts.}
    \textmd{\footnotesize The leftmost plot is for \command{nmap}{\scriptsize} and the rightmost is for \command{patator}{\scriptsize}.} Additional plots in Fig.~\ref{fig:feature_dist}.}
    \label{fig:feature-dist-bare-concap}
    \vspace{-5mm}
\end{figure}

\begin{cooltextbox}
\textsc{\textbf{Answer to RQ1.}} Our packet- and NetFlow-level analyses revealed that the network traffic generated by \concap{} resembles that of a physical network. Deviations are due to expected differences in the network channel, which are impossible to control---but do not affect the payload content.
\end{cooltextbox}

{
\subsection{Determinism of \concap{\large}'s generated traffic}
\label{ssec:deterministic}

\noindent
In our previous experiments, we found an intriguing inconsistency. Indeed, by comparing the number of packets generated via the (default) \command{nmap}{\small} and ssh-\command{patator}{\small} commands in the first experiment with those of the equivalent veriant (i.e., \command{nmap -Pn -sS -sV}{\small} and ssh-\command{patator P=1 RL=0 T=10}{\small}) of the second experiment (cf. Table~\ref{tab:broad} with Tables~\ref{tab:realistic-traffic}), we see minor deviations---despite the command and the setup being identical (for both \concap{} and the physical network). While it is known that real-world networks exhibit immense variability~\cite{apruzzese2023sok,sommer2010outside}, and hence such differences may be expected, we do not know if the same also holds for \concap{}. Therefore, we ask ourselves our second research question (RQ2):

\vspace{1mm}
{\setstretch{0.99}
\textbox{{\textsc{\textbf{Research Question \#2:}} To what extent is the traffic generated by \concap{} deterministic?}
}}

\textbf{Motivation.} There are three reasons why RQ2 is worthy of being investigated. First, to our knowledge, it is still unknown if the traffic generated by containers built on Kubernetes-related frameworks is deterministic; for instance, the authors of DetGen~\cite{clausen2019traffic} tested this hypothesis for Docker, which is different software. Second, because among \concap{}'s aims is that of fostering reproducibility (§\ref{ssec:design}), and it is hence crucial to examine the degree of similarity across multiple executions of the same scenario via \concap{}. Third, because it enables one to measure the extent to which \concap{} can approximate the ``unpredictable'' behavior of real-world networks.

\textbf{Setup.} To answer RQ2, we created four scenarios (similar to those used in §\ref{ssec:broad} and §\ref{ssec:indepth})} with increasing complexity in the traffic exchanged between the attacker and target: a simple \command{ping}{\small} scan (10 ICMP requests), a basic port scan (\command{nmap -sS}{\small}), a full port scan (\command{nmap -A -T4}{\small}), and an ssh brute-force attack (\command{patator -P=1 -RL=0 -T=10}{\small}). We capture the packets and generate the NetFlows for each scenario. We repeat each scenario 100 times. The intention is to study the degree of similarity across all of these repetitions. We also carry out the exact same operations (repeating them 10 times) on the bare-metal servers to compare with a physical setup approximating a real-world network, which should not be deterministic.

\textbf{Results.} Table~\ref{tab:concap-deterministic} reports the results (mean and std) across our trials for both setups, detailing the number of packets, byte count, and the number of NetFlows for CICFlowMeter and Argus. We analyse these results by focusing on \concap{}.
\begin{itemize}[leftmargin=*]
    \item We see no variation for the ping and basic port scan (\command{nmap -sS}{\small}) at both the packet- and NetFlow level. This shows that \textit{the network connection is reliable}: these scenarios involve sending a single request and receiving a single response (or none). Variation would only occur if the network between the attacker and target were unreliable, causing packet duplication, loss, or corruption.
    
    \item In the complex scenarios (\command{nmap -A -T4}{\small} and \command{patator -P=1 -RL=0 -T=10}{\small}), some variation is observed at the packet level, and in the case of the full port scan (\command{nmap -A -T4}{\small}), even at the NetFlow level. Similar to the realistic traffic assessment, packet-level variation arises from differences in how much data can be transmitted in a single packet and how this data is acknowledged. The flow-level variation in the full Nmap scan is caused by the aggressive mode (\command{-T4}{\small}) used by Nmap, which can overload the target.
\end{itemize}
{By comparing the results of \concap{} with those of the physical network, we see a consistent behavior at the NetFlow-level. The differences are only for the Full Port Scan and amount to less than 0.1\%, which are not significant. In contrast, more pronounced differences exist for the Full Port Scan and SSH Bruteforce attacks from a packet-level viewpoint. In these cases, \concap{} generates an average of \smamath{\approx}9\% less packets and \smamath{\approx}9\% less bytes than the physical setup.}

\begin{cooltextbox}
{\textsc{\textbf{Answer to RQ2.}} Traffic generated by \concap{} is deterministic content-wise, but the nondeterministic nature of networking results in small variations in packets and bytes.
Variations are due to how data is acknowledged (e.g., using a separate TCP packet) and should not impact the outcome of a scientific experiment. 
\concap{} does not just simulate network traffic, it generates it such that it resembles physical networks---whose real-world behavior} is unpredictable~\cite{sommer2010outside}.
\end{cooltextbox}

\begin{table}[!t]
    \caption{\textbf{RQ evaluation.} 
     We repeat the scenarios with \concap{} 100 times.} 
    \label{tab:concap-deterministic}
    \centering
    \resizebox{\columnwidth}{!}{
        \begin{tabular}{ll S[table-format = 5(2)] S[table-format = 7(2)] S[table-format = 5(2)] S[table-format = 5(2)]}
        \toprule
        \multirow{2}{*}{\textbf{Environment}} & \multirow{2}{*}{\textbf{Attack}} & \multicolumn{2}{c}{\textbf{Packets}} & \multicolumn{2}{c}{\textbf{Number of Flows}} \\
        \cmidrule{3-6}
        & & \textbf{Count} & \textbf{Sum of Bytes} & \textbf{CICFlowMeter} & \textbf{Argus} \\
        \midrule
        \multirow{4}{*}{Bare-Metal} & Ping scan & 20(0) & 1960(0) & 1(0) &  10(0) \\
        & Basic Port Scan & 13(0) & 736(0) & 5(0) &  6(0) \\
        & Full Port Scan & 2751(88) & 271551(6235) & 1091(43) &  1093(43) \\
        & SSH Bruteforce & 30935(37) & 5060797(2417) & 680(0) &  679(0) \\
        \midrule
        \multirow{4}{*}{\concap{}} & Ping scan & 20(0) & 1960(0) & 1(0) &  10(0) \\
        & Basic Port Scan & 13(0) & 694(0) & 5(0) &  6(0) \\
        & Full Port Scan & 2504(6) & 239401(4631) & 1098(1) &  1092(1) \\
        & SSH Bruteforce & 26960(354) & 4759703(23353) & 680(0) &  679(0) \\
        \bottomrule
        \end{tabular}
    }
    \vspace{-4mm}
\end{table}

{
\section{Applications: \concap{} for Research}
\label{sec:application}

\noindent
Here, we show the utility of \concap{} via proof-of-concept experiments involving noteworthy practical applications---especially from an ML viewpoint.
First, we show that \concap{} can be used to replicate prior research~(§\ref{ssec:replicability}). Then, we show that \concap{} enables creation of ML-based NIDS from scratch~(§\ref{ssec:development}). Finally, we show how to use \concap{} for assessments of existing NIDS against unseen attacks~(§\ref{ssec:enhancing}). 

\vspace{1mm}
{\setstretch{0.85}
\textbox{{\small {\textsc{\textbf{Disclaimer:}} the following ``proof of concept'' experiments are just meant to demonstrate some applications of \concap{}. We do not claim generalizable results, nor seek to cover all possible cases.}}
}}

\subsection{Replicability of Prior Research}
\label{ssec:replicability}

\noindent
Recall that our previous experiments (§\ref{sec:validation})  revealed some small differences (at least at the packet level) in the data generated by \concap{} w.r.t. of a physical network. Hence, we ask ourselves: are such differences ``significant'' from the viewpoint of ML-based network-intrusion detection? 

\textbf{Goal.} We explore this question by attempting to reproduce the results of prior work focusing on ML-based NIDS. Specifically, we consider the experiments carried out by Engelen et al.~\cite{engelen2021troubleshooting} and Liu et al.~\cite{liu2022error}, which involved assessment of supervised ML-based classifiers analysing network flows created via a (fixed) version of CICFlowMeter (the same one we used for \concap{}). Specifically, the NetFlows pertain to the well-known CICIDS17 and CICIDS18 benchmark datasets~\cite{sharafaldin2018toward}; such NetFlows have been rigorously labeled by the authors of~\cite{engelen2021troubleshooting,liu2022error}. Both CICIDS17 and CICIDS18 include, among others, network-traffic data generated with ssh-\command{patator}{\small} (which we used in §\ref{sec:validation}). Our intention is to train various ML models on the NetFlows of ssh-\command{patator}{\small} included in CICIDS17 and CICIDS18, and then testing such classifiers on the NetFlows of ssh-\command{patator}{\small} generated via \concap{}, and then comparing the results: our expectation is that the classifiers trained on either ``version'' (i.e., CICIDS17/18, or \concap{}) of ssh-\command{patator}{\small} shall always be able to detect the other version---thereby demonstrating that the data generated by \concap{} is functionally equivalent (ML-wise) to that of prior research.

\textbf{Setup.}
We follow the instructions provided in~\cite{liu2022error,engelen2021troubleshooting}, and retrieve the (fixed) version of CICIDS17 and CICIDS18 (more details in Appendix~\ref{sapp:benchmark}). Then, for each dataset, we partition the NetFlows according to their classes. For this experiment, we are only interested in the ssh-\command{patator}{\small} class (which, in both CICIDS17 and CICIDS18, was launched with the default options), which we denote as \smacal{P}; and the benign class, which we denote as \smacal{B}. Then, we take the NetFlows generated by \concap{} with the default ssh-\command{patator}{\small} (see §\ref{ssec:broad}), which we denote as \smacal{\overbar{P}}. For a broad assessment, we consider five different ML-based classifiers: Random Forest (RF), Decision Tree (DT), XGBoost (XGB), Support Vector Machine (SVM), as well as a Deep Neural Network (DNN); these classifiers have been tested in~\cite{liu2022error,engelen2021troubleshooting}, typically obtaining near-perfect performance. Then, for each dataset, we train each classifier on 80\% of \smacal{P} and 80\% of \smacal{B}, and test it on the remaining 20\% of \smacal{B} (for the false positive rate, \smamath{fpr}), the remaining 20\% of \smacal{P}, and all of \smacal{\overbar{P}} (computing the true-positive rate, \smamath{tpr} for both sets). We then repeat this process, but by switching \smacal{P} with \smacal{\overbar{P}}. We repeat this procedure 5 times for statistical robustness. Note that such a workflow complies with the best practices of ML-based assessments in cybersecurity and ML-NIDS~\cite{arp2022and,apruzzese2023sok}. The experimental source-code is in our repository~\cite{repository}.

\begin{table}[t]
    \caption{\textbf{Reproducibility of prior work.} We show \concap{}-generated data is functionally equivalent to that of existing benchmark datasets.} 
    \label{tab:demo_benchmark}
    \vspace{-2mm}
    \centering
    \resizebox{0.8\columnwidth}{!}{
    {
        \begin{tabular}{
        c l ?
        r r r ?
        r c c 
        }
        \toprule
        & Train Set & \multicolumn{3}{c?}{\smacal{B}+\smacal{P}} &  \multicolumn{3}{c}{\smacal{B}+\smacal{\overbar{P}}}  \\
        & Test Set & \smacal{B} & \smacal{P} & \smacal{\overbar{P}} & \smacal{B} & \smacal{P} & \smacal{\overbar{P}} \\
        \midrule
        \multirow{5}{*}{\rotatebox[origin=c]{90}{CICIDS17}} & DT & $<$0.001 & 0.997 & 0.917 &  $<$0.001 & 0.980 & 1.000 \\
        & RF & 0.000 & 0.998 & 1.000 & 0.000  & 0.986 & 1.000 \\
        & XGB & $<$0.001 & 0.998 & 0.869  & $<$0.001  & 1.000 & 1.000 \\
        & SVM & $<$0.001 & 0.993 & 0.907 & $<$0.001  & 0.993 & 1.000 \\
        & DNN & 0.000 & 0.998 & 0.920 & $<$0.001  & 0.996 & 1.000 \\
        \midrule
        \multirow{5}{*}{\rotatebox[origin=c]{90}{CICIDS18}} & DT & 0.000 & 1.000 & 0.859 &  $<$0.001 & 1.000 & 1.000 \\
        & RF & 0.000 & 1.000 & 0.859 & $<$0.001  & 1.000 & 1.000 \\
        & XGB & 0.000 & 1.000 & 0.859  & $<$0.001 & 0.984 & 1.000 \\
        & SVM & 0.000 & 1.000 & 0.907 & $<$0.001 & 1.000 & 1.000 \\
        & DNN & 0.000 & 1.000 & 0.907 & $<$0.001 & 1.000 & 1.000 \\

        \bottomrule
        \end{tabular}
    }}
    \vspace{-6mm}
\end{table}

\textbf{Results.} We report the results of our tests in Table~\ref{tab:demo_benchmark}. The table shows the \smamath{fpr}, which is always less than 0.001 (a result consistent with prior work~\cite{liu2022error,engelen2021troubleshooting}), even when the classifiers were trained with \concap{}'s generated data. Then, looking at the \smamath{tpr}, we can see that the results on \smacal{\overbar{P}} align with those on \smacal{P}, both on CICIDS17 and CICIDS18, and irrespective of whether the training was done on 80\% of \smacal{\overbar{P}} or \smacal{P}. Indeed, despite not being ``perfectly'' aligned\footnote{{We conjecture that such minor difference may stem---besides from unpredictable network effects---from potential labeling inaccuracies in the (fixed) CICIDS17 or CICIDS18, since these tasks have been done manually (whereas \concap{} does so automatically); they can also be due to different machines (the creators of CICIDS17/18 do not provide low-level hardware details~\cite{sharafaldin2018toward}).}} the \smamath{tpr} always shows that these attacks can be detected, since the \smamath{tpr} is always above 0.85. Importantly, this result is shared across all models/classifiers. Thus, we can say that, from an ML viewpoint, and at least according to these experiments, the (labeled) NetFlows generated via \concap{} can be used to replicate prior research and derive the same conclusions.

\subsection{Development of ML-based NIDS for the Real World}
\label{ssec:development}

\noindent
We consider the case in which a researcher may want to develop an ML-based NIDS \textit{from scratch}. We assume that the researcher has access to a real-world network and that the hosts of such a network are not compromised. In this context, we wonder: can \concap{} be used to develop an ML-NIDS that detects malicious traffic generated by real-world networks?

\textbf{Goal.} This experiment is conceptually similar to the one discussed in the previous subsection (§\ref{ssec:replicability}). At a high level, we will use \concap{} to generate malicious network activities, and do the same on a real-world setup; and then see if using the malicious NetFlows generated via \concap{} to train an ML-based classifier yields a model that can detect the real-world version of the malicious network activities---and vice versa. The difference, however, is that we are not relying on benchmark datasets here: this experiment is carried out on a real-world ``smart home'' network.\footnote{{Yes, we captured traffic from a network regularly used by real people, and infected hosts deployed in such a network. We obtained permissions by the users/residents. We discuss ethical considerations at the end of the paper.}} Our intention is twofold: first, verify if the results we achieved in our previous experiment also hold here---thereby reinforcing the claim that the data generated by \concap{} is unlikely to alter the conclusions of a ML-based experiment; second, examine how well the model trained with \concap{}'s NetFlows can detect their real-world variant (which technically represents the ``true attack'' that the ML-NIDS should protect against). 

\textbf{Setup.} 
For this experiment, we consider a real, physical network encompassing \smamath{\approx}50 active hosts (note: this network is different from the one of the ``bare-metal'' servers used in §\ref{sec:validation}). These hosts include a mix of IoT devices, gaming consoles, laptops, as well as smartphones. We captured \smamath{\approx}20GB of network traffic in this network, which we verified to be clean of malicious activities. To capture the malicious activities, we consider two hosts (i.e., two laptops) deployed in this network, one acting as the ``target'' and the other as the ``attacker''. To align this experiment with the one in §\ref{ssec:replicability}, we consider the ssh-\command{patator}{\small} as our attack. Hence, we deploy an SSH server on the ``target'' host, and use the ``attacker'' host to launch ssh-\command{patator}{\small} (with its default options) against the SSH server. We captured the corresponding PCAP trace (on the ``target'' host), extracted the NetFlows (via CICFLowMeter) and manually labeled them. We then follow the same workflow as in §\ref{ssec:replicability} for the ML-based experiments, using the same notation. These experiments are also available in our repository~\cite{repository} and more details on the overarching real-world network and packet capture are provided in the Appendix~\ref{sapp:homenetwork}.

\textbf{Results.} The results of this experiment are in Table~\ref{tab:demo_real}. We can see a similar trend to that shown in the previous experiment (cf. Table~\ref{tab:demo_benchmark} with Table~\ref{tab:demo_real}). In particular, the models trained with \concap{}-generated data (\smacal{B}+\smacal{\overbar{P}} columns) have a near-zero \smamath{fpr} and a near-perfect \smamath{tpr}, both on the malicious NetFlows generated via \concap{} (i.e., \smacal{\overbar{P}}), which is expected; and on those generated by the physical hosts (i.e., \smacal{P}). Moreover, such a good performance also encompasses the case wherein the models are trained only with real-world data (\smacal{B}+\smacal{P} columns).
Altogether, these finding further confirm that \concap{} can be used to carry out ML- and NetFlow-based experiments (thereby reinforcing our conclusions related to §\ref{ssec:replicability}); and also show that \concap{} can be used to develop ML-based NIDS meant to be deployed in the real world.

\begin{table}[t]
    \caption{\textbf{Real-world equivalence.} Network traffic generated with \concap{} is compatible with a real-world ''smart home'' network.} 
    \label{tab:demo_real}
    \vspace{-2mm}
    \centering
    \resizebox{0.8\columnwidth}{!}{
    {
        \begin{tabular}{
        l ?
        r r r ?
        r r  r
        }
        \toprule
        Train Set & \multicolumn{3}{c?}{\smacal{B}+\smacal{P}} &  \multicolumn{3}{c}{\smacal{B}+\smacal{\overbar{P}}}  \\
        Test Set & \smacal{B} & \smacal{P} & \smacal{\overbar{P}} & \smacal{B} & \smacal{P} & \smacal{\overbar{P}} \\
        \midrule
        DT & $<$0.001 & 1.000 & 0.899 &  $<$0.001 & $>$0.999 & $>$0.999 \\
        RF & 0.000 & 1.000 & 0.899 & 0.000 & 1.000 & 1.000 \\
        XGB & 0.000 & $>$0.999 & 0.889  & $<$0.001 & 1.000 & 1.000 \\
        SVM & 0.000 & 1.000 & 1.000 & $<$0.001 & 1.000 & 1.000 \\
        DNN & $<$0.001 & 1.000 & 1.000 & $<$0.001 & 1.000 & 1.000 \\

        \bottomrule
        \end{tabular}
    }}
    \vspace{-5mm}
\end{table}

\subsection{Security Assessment of Existing (ML-based) NIDS}
\label{ssec:enhancing}
\noindent
Our previous experiments showed that \concap{}'s generated data can be used to develop ML-based NIDS, showing that, by training classifiers on some malicious data of a ``known'' attack, it is possible to detect future instances of the same attack. Here, we show how \concap{} can be used to produce new knowledge. Specifically, we ask ourselves: how can \concap{} be used to test existing NIDS against ``unseen'' attacks?

\textbf{Threat Model.} 
We consider a defender that uses an ML-NIDS developed by using either the CICIDS17 or CICIDS18 dataset (in their fixed version~\cite{engelen2021troubleshooting,liu2022error}); such an assumption serves for broad coverage of diverse use cases, since these datasets are well-known ``benchmarks'', encompass a variety of attacks, and are used even in recent and top-tier publications (see, e.g.,~\cite{flood2024bad}). Differently from our previous experiments (in §\ref{ssec:replicability}) here we use all of the malicious data (not just that of ssh-\command{patator}{\small}) in these datasets to develop our models, leading to much larger, and therefore more effective, detectors. The attacker seeks to evade the detection by using recent exploits, whose corresponding network traffic has not been included in the training dataset of the detector used by the defender. Without loss of generality, we consider three real and recently published exploits, taken from the common vulnerabilities and exposures (CVE) database: CVE-2024-47177~\cite{cve1} (related to OpenPrinting Cups), CVE-2024-36401~\cite{cve2}  (related to GeoServer), CVE-2024-2961~\cite{cve3}  (related to GNU C Library). The attacker hence exploits these CVEs, and hopes that such malicious activities are not detected by the defender.

\textbf{Approach.}
We train five ML-based classifiers (RF, DT, SVM, DNN, XGB) on the whole CICIDS17 and CICIDS18 datasets; we confirm that such ML-NIDS obtain the same performance as prior work~\cite{engelen2021troubleshooting, liu2022error}. Then, we use \concap{} to generate network-traffic data of the three considered CVEs. We setup \concap{} accordingly, so that the target host can be ``exploited'' via the commands included in the CVE and executed on the attacker host.
To expand our coverage (and also to show the capabilities of \concap{}), we perform the captures by varying the overarching network conditions, specifying different values for \textit{delay}, \textit{loss}, \textit{corrupt}, and \textit{duplicate}, totaling over 320 combinations. Indeed, from a practical viewpoint, \concap{} makes such repetitions trivial to carry out, and it is sensible to assume that an attacker may exploit any given CVE randomly, i.e., when the network conditions may be subject to bursts or noise. Overall, we capture 350k packets, corresponding to 4800, 640, and 1280 NetFlows for CVE-2024-47177, CVE-2024-36401, CVE-2024-2961, respectively. We then test our 10 models (5 per dataset) on these NetFlows. 

\textbf{Results.}
We report the results in Table~\ref{tab:new_data} (Appendix~\ref{app:results}). 
Our models are unable to reliably recognize the CVE NetFlows as malicious (the \smamath{tpr} is always below 0.6, and below 0.25 in most cases). 
Such an insight can be used as a signal that the ML-NIDS must be retrained---which is a trivial task since it is simply necessary to enhance the training dataset with the data generated by \concap{} and re-train the ML models by, e.g., leveraging the well-known adversarial training technique~\cite{apruzzese2023sok}. (In Appendix~\ref{sapp:multistep}, we also demonstrate a full-fledged multi-step and multi-host attack by using \concap{}.)

}
\section{Discussion and Critical Analyses}
\label{sec:discussion}
\noindent
{We evaluate the computing requirements needed to run \concap{}~(§\ref{ssec:runtime}), discuss the limitations of our research~(§\ref{ssec:limitations}), and draw lessons learned from our experiments (§\ref{ssec:lessons}).

\subsection{Computing Requirements and Operational Runtime}
\label{ssec:runtime}

\noindent
We show that \concap{} can seamlessly run also on commodity hardware (e.g., laptops) by measuring its operational runtime.

\textbf{Approach.} We test \concap{} by performing various experiments on two setups: a multi-node Kubernetes cluster (the same used in §\ref{sec:validation}), and a laptop with an Intel i7-1265CPU and 23GB RAM (more hardware details in the Appendix~\ref{sapp:runtime}). Specifically, we consider the 12 combinations of \command{nmap}{\small} (also used in §\ref{ssec:indepth}). Note that what we are interested in is the background utilization of resources (CPU and RAM) as well as the overall initialization time, i.e., the time required before the ``attacker'' host begins issuing its commands. Anything beyond this step does not depend on \concap{} (e.g., if the attacker wants to DDoS the target, or if the scenario requires setting up a target running resource-intensive services, then such requirements are unrelated to \concap{}'s resource utilization).

\textbf{Results.} We report the results in Table~\ref{table:concap-perf} (in the Appendix~\ref{sapp:runtime}, which also provides more details). On average, less than 3 seconds elapse (on both setups) before the \command{nmap}{\small} command is launched. We also see that, on both setups, the containers running \command{nmap}{\small} requires less than 2MiB of RAM (across all of our experiments); whereas \concap{}'s memory footprint is less than 40MB. In terms of CPU, the utilization varies (which is expected, since it depends also on the dynamic allocation of the machine) but, on average, the max-CPU used was 10\% for the cluster and 6\% for the laptop. We can hence conclude that \concap{} is also intrinsically lightweight.

}

\subsection{Scope and Limitations}
\label{ssec:limitations}

\noindent
We advance the state of the art by {providing a tool, \concap{}, to foster future research on NID. Our contribution are aimed at \textit{research}: real-world deployments are outside our scope~(as we clearly remarked in §\ref{sec:background}). Although we did rely on real-world assessments, such assessments primarily served to prove the utility of \concap{} for this specific purpose. 

As for limitations, we acknowledged \concap{} is not fully deterministic: while this property can be a strength (e.g., by running the same scenario multiple times, it is possible to collect more ``realistic'' data) it can also be a weakness (e.g., for reproducibility of some experiments expecting to achieve perfect matches). However, such a limitation is expected due to the non-deterministic nature of modern networks/protocols (see §\ref{sssec:packetlevel}). Second, \concap{}'s labeling is addressed to network flows, and cannot hence be used for labeling of other datatypes which can be used for NIDS purposes (e.g.,~\cite{venturi2023arganids}; yet, as we argued (in §\ref{ssec:design}) the majority of papers on NIDS rely on network flows for their analyses. 
Third, and related to the experiments carried out in this work, we believe the answer to our major research questions (in §\ref{sec:validation}) to be correct and also generalizable (for instance, even the authors of DetGen~\cite{clausen2019traffic} obtained similar results to ours, albeit with different software). In contrast, we acknowledge that the experiments in §\ref{sec:application} cannot be used to cover all possible use-cases: this is why we emphasized that the experiments in that section serve as a ``proof of concept'' to show potential applications of \concap{}. 

Finally, we emphasize that \concap{} can only generate data according to the specifications provided by the developer.

\subsection{Lessons Learned}
\label{ssec:lessons}

\noindent
Let us summarize the major takeaways of our work.

First, \concap{} automatically generates network-traffic data conforming to ``any'' attack. Such data resembles that of a physical network, including its non-deterministic nature~(§\ref{sec:validation}). We \textit{manually verified all data labeled} by \concap{}, and the labeling was always correct: the NetFlows always pertained to the malicious commands specified in the configuration file. Such a verification is confirmed by our experiment showing that such data is functionally equivalent to that of benchmarks/real-world networks (§\ref{ssec:replicability} and §\ref{ssec:development}).

Given the above, \concap{} can be used in a variety of ways (some of which shown in §\ref{sec:application}). For instance, through \concap{}, it is possible to produce new knowledge (or test new hypotheses) without the need (and risk) to carry out experiments entailing malicious activities in real-world networks. Crucially, and as a byproduct of the experiments in §\ref{ssec:enhancing}, \concap{} enables early-assessments of existing NIDS (not-necessarily reliant on ML methods) against recent threats. For instance, through \concap{}, the owners of an NIDS can safely test their systems by {\small \textit{(i)}}~taking any new CVE and {\small \textit{(ii)}}~running it on \concap{}, and then {\small \textit{(iii)}}~either replay the PCAP trace or submit the NetFlows to the NIDS. Depending on the outcome, it is then possible to \textit{enhance} the NIDS by, e.g., using adversarial training~\cite{apruzzese2023sok} with \concap{}'s generated data. 

Finally, through \concap{}, future research does not need to ``share data'' (which can require abundant storage space), rather ``sharing the scenario configuration file'' (typically less than 1KB in size) and the experimental details (e.g., the CVE used) is sufficient to reproduce prior results. 

\vspace{1mm}
{\setstretch{0.9}
\textbox{{\small {\textsc{\textbf{Q\&A:}} We clarify some remarks (e.g., fitness to SaTML and realism of \concap{}'s data) made by reviewers in the Appendix~\ref{sapp:concerns}}}
}}

}

{
\section{Related Work}
\label{ssec:related}
\noindent
We are not aware of any open-source tool that fulfills the same goals as \concap{}. Here, we first summarize related literature (§\ref{ssec:summary}), then compare \concap{} with the two closest works we could find (§\ref{ssec:comparisons}), and finally suggest avenues in which \concap{} can benefit future related research (§\ref{ssec:future}).

\subsection{Literature Summary}
\label{ssec:summary}
\noindent
Some orthogonal works propose testbeds that do not provide the functionalities of \concap{}: e.g., Gotham~\cite{saez2023gotham} cannot ensure fine-grained labeling of malicious datapoints, whereas I2DT~\cite{cordero2021generating} can only inject packets in a PCAP.

Closely related works are DetGen~\cite{clausen2019traffic} and SOCBED~\cite{uetz2021reproducible}; we tried to find more works on ``generation of realistic network traffic'' by looking at the accepted papers in various top-tier conferences (NDSS, IEEE S\&P, USENIX Sec, ACM CCS) since 2019 (similarly to~\cite{arp2022and, apruzzese2023real}). We found that the only related work was netUnicorn~\cite{beltiukov2023search}. Then, we expanded our search via snowballing~\cite{wohlin2014guidelines} and (recursively) looked for all peer-reviewed papers cited by~\cite{clausen2019traffic,beltiukov2023search}, and identified four related works~\cite{beltiukov2023pinot,khan2024harnessing,buhler2022generating, zhoutrafficformer, guthula2023netfound}. Let us position  \concap{} within these related works. 

PINOT~\cite{beltiukov2023pinot} and netUnicorn~\cite{beltiukov2023search}, can generate benign traffic but raise security risks for generating malicious traffic (see §\ref{ssec:obtaining}). 
To generate benign traffic at scale, netMosaic~\cite{khan2024harnessing, buhler2022generating} harnesses public code repositories to automatically capture network traffic for a wide range of applications, but does not allow fine-grained labeling of malicious datapoints. 
Finally, Zhou et al.~\cite{zhoutrafficformer} propose to use foundational models to artificially ``augment'' any given network-traffic dataset; however, it is unclear if such methods can produce data resembling a physical setup (there is no real-world assessment in~\cite{zhoutrafficformer}). 

\subsection{Low-level Comparisons}
\label{ssec:comparisons}
\noindent
We compare \concap{} with DetGen~\cite{detgenGithub} and SOCBED~\cite{socbedGithub}.

\textbf{\concap{} vs DetGen.}
First, DetGen~\cite{clausen2019traffic} does not allow parallelism by design. Second, DetGen does not generate and label the NetFlows of any given experiment, thereby forcing the developer to do so manually---which is error-prone~\cite{engelen2021troubleshooting}. Finally, DetGen does not allow the same degree of flexibility provided by \concap{}. To provide evidence of this claim, we looked at the public repository of DetGen (available at~\cite{detgenGithub}), inspected its source code, and observed the following: 
{\small \textit{(i)}}~we did not find any way to set the available CPU/RAM of the attacker/target; 
{\small \textit{(ii)}}~a predefined timeout is required for every scenario, potentially stopping the scenario before successful completion; 
{\small \textit{(iii)}}~comments in the code state that the attack sometimes starts before traffic is captured, highlighting the lack of fine-grained control over the scenario execution by DetGen.
Although we tried to compare the traffic generated by both approaches, we were unable to run any of their example scenarios (due to deprecated software/runtime errors that we could not troubleshoot with the provided documentation).

\textbf{\concap{} vs SOCBED.}
First, SOCBED primarily targets log-data collection: even though it can create PCAPs, it does not {\small \textit{(a)}}~extract NetFlows by default, or {\small \textit{(b)}}~label them. The latter is crucial: manual post-hoc labeling of attack traffic is unreliable due to the background noise generated by the concurrent simulation of benign and malicious behaviors. Second, SOCBED relies on a \textit{fixed} network topology designed for long-running experiments running \textit{virtual machines} on a single host. Their sample scenario requires around one hour to complete, including a 15-minute setup time. In contrast, \concap{} supports fast (startup time of seconds), parallel execution of isolated attack scenarios across multiple hosts. 
Finally, we found no example to configure network characteristics, such as per-host bandwidth or delay, in SOCBED.

\vspace{1mm}
{\setstretch{0.8}
\textbox{{\small {\textsc{\textbf{Replication:}} We show in the Appendix~\ref{sapp:DetGen} and~\ref{sapp:socbed} how to replicate one of DetGen's and SOCBED scenarios with \concap{}.}
}}
}

\subsection{Future Work}
\label{ssec:future}

\noindent
Future research can use \concap{} to investigate open problems in NIDS~\cite{ceschin2024machine,cina2024machine}, such as: robustness to concept drift~\cite{andresini2021insomnia, wang2022enidrift}, explainability~\cite{bhusal2023sok, mink2023everybody,wei2023xnids}, false alarms~\cite{alahmadi202299, van2022deepcase,vermeer2023alert,fu2023point}, or development of novel detection techniques (not necessarily relying on NetFlows, such as~\cite{mirsky2018kitsune, sharif2024drsec, rehman2024flash, yang2023prographer}). \concap{} can technically also be used for generation of \textit{benign} labeled traffic (useful for, e.g., traffic classification~\cite{azab2024network,cebere2024understanding,siracusano2022re,rimmer2022trace, schafer2022accurate}). To further demonstrate the broad applicability of \concap{}, we use its data to test other detection approaches~\cite{wei2023xnids,luay2025multimodal}, based on deep learning or transformers, in Appendix~\ref{sapp:contemporary}. 
\section{Conclusions}
\label{sec:conclusions}

\noindent
Our paper is a stepping stone for future research in NID.

With \concap{}, we enable researchers---especially those without access to real-world networks—--to conduct realistic security assessments in networking contexts. \concap{} automatic (and correct) labeling capabilities remove the burden of carrying out manual annotation. Moreover, \concap{} open-source imprint fosters reproducibility and transparency.

\section*{Acknowledgments}
The authors would like to thank the anonymous SaTML'26 reviewers for the feedback received, which substantially improved our work. Part of this research was funded by the Hilti Corporation. This research was also partially supported by the BOSA-AIDE project funded by the Belgian SPF BOSA with reference number 06.40.32.33.00.10. This work originated from a research stay at the University of Liechtenstein and was supported by the Research Foundation -- Flanders (FWO) under grant number V450224N.

\bibliographystyle{IEEEtran}

{\footnotesize

}

\appendices
\section{Ethics, Open Science, and Clarifications}
\label{app:ethics}

\noindent
Following the best scientific practices, and to also comply with the CfP, in this appendix we: provide some ethical comments (Appendix~\ref{ethics}); describe the technical content we will publicly release (Appendix~\ref{artifact}); and elaborate our response to some valid remarks that we received in previous versions of this paper (Appendix~\ref{sapp:concerns}), while also disclosing our ``responsible usage'' of compute.

\subsection{Ethical Remarks}
\label{ethics}

\noindent
We make the following ethical considerations.

\textbf{Prevention of harm.}
For our research, we carried out experiments entailing ``network attacks'' (i.e., ssh bruteforcing~\cite{patator}) on a real-world network. The authors received permission to do so, and no software/hardware was damaged as a result of these operations.

\textbf{Data Confidentiality.}
For our research, we have collected data from two physical networks: that of the bare-metal setup, and the real-world one encompassing \smamath{\sim}50 hosts. We will release the full data (PCAP and NetFlows) for the bare-metal setup. For the other network, we will only provide the (labeled) NetFlows, which do not leak private data (we will anonymise the IP addresses and ports). We received permission by the owners to share this data publicly.

\textbf{Respectfulness}. 
We emphasize that our intention is not to ``point the finger'' against any prior work. We found a bug in the current implementation of DetGen~\cite{detgenGithub} that prevented us from proceeding in our comparison (see §\ref{ssec:related}). We contacted the developers of DetGen, informing them of the issue. They responded and confirmed that the current code is outdated and does not function properly. We are currently corresponding with them, offering our help to fix the problem.

\subsection{Data Availability}
\label{artifact}

\noindent
All necessary artifacts to replicate the results in the paper will be released together with \concap{} allowing researchers to create their own traffic generation experiments. Specifically, all our resources are available at: \url{https://github.com/idlab-discover/ConCap}~\cite{repository} (an anonymised version of this repository was also provided for peer review). The repository contains the following resources:

\begin{itemize}[leftmargin=*]
    \item The source code of \concap{} (which is based on open-source libraries)

    \item The containers used for our experiments.

    \item The PCAP and (labeled) NetFlows generated by \concap{} that we used for our experiments.

    \item The PCAP trace and NetFlows of the bare-metal setup.

    \item The labeled NetFlows of our real-world network of 50 hosts; (we cannot release the PCAP for privacy). 

    \item The notebooks of our experiments (for reproducibility).

    \item The additional data (PCAP and labeled NetFlows) used to showcase that \concap{} can be used to generate ``new'' data conforming to recent attacks.
\end{itemize}
We have the permission to share all of the above.

\subsection{Clarifications on our Research}
\label{sapp:concerns}
\noindent
We provide additional clarifications on four aspects: the labeling accuracy of \concap{}, the realism of \concap{}'s generated data, the necessity to provide ad-hoc configuration files to run \concap{}, the resources used to carry out our experimental evaluation, and fitness of our work to the SaTML community. To facilitate understanding, we organize this appendix in a ``Question-and-Answer'' format.

\textbf{How does \concap{} ensure that the NetFlows are correctly labeled?} This is clarified in Section~\ref{ssec:principles}. The major difficulty in labeling network data (and, in our case, NetFlows) is that it is difficult precisely distinguish malicious datapoints among the myriad of activities carried out by modern machines. For instance, the labeling of most popular benchmarks is done by {\small \textit{(i)}}~generating some network traffic in synthetic environments, {\small \textit{(ii)}}~creating the corresponding NetFlows, and {\small \textit{(iii)}}~apply coarse labeling strategies---such as ``assign the label \$maliciousLabel to all NetFlows generated by \$IPaddress between \$startTime and \$endTime using ports \$dPort and \$sPort'' (see~\cite{flood2024bad}). Such strategies may inevitably lead to labeling as malicious also benign/background traffic, and they may also underestimate the actual malicious traffic (e.g., there may be malicious activities related to a given malware that entail ports different from \$dPort or \$sPort''). In contrast, the labeling applied by \concap{} \textit{does not have such a problem by design}: thanks to its isolated environment and to the fact that the ``attacker'' host runs only the (supposedly malicious) commands specified in the configuration file, it is guaranteed that the packets generated by the attacker host will include only the network activities that pertain to the provided command(s). In our experiments, we have verified that this holds true (see §\ref{ssec:lessons}); moreover, the fact that the \concap{}'s generated-and-labeled data yielded ML-NIDS that have the same performance as in prior work (see experiment in §\ref{ssec:replicability}) further confirms that \concap{} provides accurate labeling---by design and without requiring human intervention.

\textbf{\concap{} generates network traffic in a synthetic environment. How can such traffic be realistic?} From a technical viewpoint, network packets have no notion of ``real-world'' or ``synthetic'' environment. Therefore, the payload (i.e., the actual data that flows over a network and that is exchanged between two endpoints) should be identical whether the traffic capture occurs in a ``synthetic'' or ``physical'' testbed. However, networks---irrespective of ``where'' they are---are characterized by having immense variability~\cite{sommer2010outside}. For instance, two organizations having the exact same hosts which carry out the exact same activities can have different network traffic data because, e.g., they may have different bandwidth. Thanks to \concap{}'s configuration file, it is possible to specify parameters of the network channel that can lead to a better approximation of the real-world network behavior. Note, however, that from the viewpoint of the malicious payload, we have empirically verified (in §\ref{sec:validation}) that there are no differences.

\textbf{\concap{} requires manual effort to define the configuration files. Wouldn't this offset the benefit of providing automatic traffic generation and labeling?} We respectfully disagree. The analysis done by Flood et al.~\cite{flood2024bad} (and, previously, also by Liu et al.~\cite{liu2022error} and by Engelen et al.~\cite{engelen2021troubleshooting}) show that manual labeling requires extensive effort. In contrast, with \concap{}, it is just necessary to, e.g., read the documentation of a given CVE and setup the hosts accordingly to produce a ``configuration file'' that can be used as a blueprint to carry out essentially endless variants of a given malicious activity. For instance, for the attacks we captured in §\ref{ssec:enhancing} (and also in Appendix~\ref{sapp:multistep}), we only took \smamath{\approx}2 hours (we did them in the timespan between a ``rebuttal phase'' of a security conference, and we can provide evidence of this). Such effort is objectively smaller than that required to, e.g., set up a full-fledged network environment (even via virtual machines), capture the traffic, generate the NetFlows, and precisely label such NetFlows.

\textbf{What is the memory footprint of your research?} Our experiments did not use a lot of compute (we report the training times of our models in Appendix~\ref{app:results}) and \concap{} requires a negligible amount of computing resources to run (see §\ref{ssec:runtime}). We do not believe more experiments are necessary to prove any of our major claims: doing so (e.g., to show how LLMs can benefit from \concap{}'s data) would be a waste of resources for the purpose of this specific paper. 

\textbf{Can you clarify how \concap{} represents a significant contribution to the SaTML research community?}
First, we acknowledge that our paper's primary contribution, \concap{}, is primarily an engineering effort. However, as shown by the EuroS\&P'24 Best Paper Award~\cite{flood2024bad} (and also by other recent works, e.g., ACSAC'25~\cite{wangr2025rr}), the ML-NIDS research community is in desperate need of ``better datasets''. Our contribution, \concap{}, is our response to such a need, given that it solves the labeling issue while generating valid data for ML-NIDS research experiments. Moreover, we stress that the development of \concap{} goes beyond simply deploying containers. Configuring and executing reproducible, labeled network experiments end-to-end---especially in the context of \textit{security} research---is not supported by Kubernetes out of the box. For instance, Kubernetes has plugins for networking of containers, but none provides support for traffic shaping (e.g., controlling bandwidth or latency). Moreover, Kubernetes does not natively support packet capture, NetFlow generation, or any form of labeling---the latter being the major problem affecting existing NIDS datasets~\cite{flood2024bad}. While some of these steps may seem trivial in isolation, executing them reliably and reproducibly across scenarios is a known challenge.
To sum up, \concap{} is a tool to ``unlock'' new research to address the many open problems in the ML-NIDS domain: without ``trustworthy'' (which we intend as ``correctly-labeled'') data, it is difficult to provide convincing solutions. \concap{} therefore enables \textit{trustworthy data curation} in a way that is \textit{safe to practically use} (since the capture occurs in a safe setup), both of which being themes that align with SaTML's vision.
\section{\concap{\huge} Configuration}
\label{app:concap-configuration}

\noindent
We provide information for practical use of \concap{}. First, the full configuration files for a \concap{} scenario and NetFlow extractor are given with all the possible configuration options. Then, a step-by-step guide is provided to replicate a scenario from DetGen~\cite{clausen2019traffic}.

\begin{lstlisting}[
frame=single,
breaklines=true, 
basicstyle=\ttfamily\footnotesize,
caption=A \concap{} scenario configuration file describing a full port scan via nmap against an Apache webserver.,
label={lst:scenario-file},
basicstyle=\scriptsize,
belowskip=-5mm,
float=t
]
attacker:
  name: nmap
  image: instrumentisto/nmap:latest
  atkCommand: nmap $TARGET_IP -A -T4
  atkTime: 30s
  cpuRequest: 100m
  memRequest: 100Mi
  cpuLimit: 500m
  memLimit: 500Mi
target:
  name: httpd
  image: httpd:latest
  filter: host $ATTACKER_IP and host $TARGET_IP and not arp
  cpuRequest: 100m
  memRequest: 100Mi
  cpuLimit: 500m
  memLimit: 500Mi
network:
  bandwidth: 100mbit
  queueSize: 100ms
  limit: 10000
  delay: 0ms
  jitter: 0ms
  distribution: normal
  loss: 0%
  corrupt: 0%
  duplicate: 0%
  seed: 0
labels:
  label: 1
  category: port-scan
  subcategory: nmap
  scenario: nmap_A_T4
  
\end{lstlisting}

\subsection{NetFlow Exporter Configuration}
\label{sapp:processing-config}

\noindent
Automatic NetFlow extraction in \concap{} is set-up by processing configuration files. A flow extractor is created for each file, which exports NetFlows from the scenario's network traces. An example configuration file in Listing~\ref{lst:processing-file} for \textit{Argus} has 3 configuration options: a ``name'', ``containerImage'', and ``command''. The name and container image are used to deploy the NetFlow exporter container. The command is responsible for processing the network capture file and outputting the extracted network flows as a CSV file.

\begin{lstlisting}[
frame=single,
breaklines=true, 
basicstyle=\ttfamily\footnotesize,
caption=Argus Processing Definition for ConCap,
label={lst:processing-file},
basicstyle=\footnotesize,
float=t
]
name: argus
containerImage: ghcr.io/idlab-discover/concap/argus:latest
command: "argus -r $INPUT_FILE -S 60s -w - | ra -r - -c, > $OUTPUT_FILE"
\end{lstlisting}

\subsection{Configuration of a DetGen Scenario in \concap{\large}}
\label{sapp:DetGen}
\noindent
To demonstrate the flexibility of \concap{}, we have created a step-by-step guide to implement one of the predefined scenarios from DetGen~\cite{clausen2019traffic}. We selected the well-documented ``capture-020-nginx'' scenario, which uses ``siege'', an HTTP load testing and benchmarking tool, to target the ``nginx'' HTTP server and reverse proxy. In this guide, we show how to replicate the ``capture-020-nginx'' scenario in \concap{} using processing pods and scenario definitions.

\textbf{NetFlow Configuration}
DetGen does not support automated NetFlow generation, thus we skip the processing pods.

\textbf{Scenario Configuration}
The \concap{} scenario is detailed in Listing~\ref{lst:replicated-detgen}. Here, the attacker is configured to use \textit{siege} for 10 seconds with 10 simulated users, each making 1,000 requests to the reverse proxy's index page. The IP address of the target is assigned through environment variable expansion. Since there is no official Docker image for \textit{siege}, we could have opted for one of the many community-built images. However, to demonstrate the simplicity of creating a custom image, we provide a minimal Dockerfile in Listing~\ref{lst:siege-image}, which we then push to our GitHub Container Registry. For the target, we use the official \textit{nginx} DockerHub image. To execute the scenario twice, we duplicate the scenario definition file.

\begin{lstlisting}[
frame=single,
breaklines=true, 
basicstyle=\ttfamily\footnotesize,
caption=The replicated capture-020-nginx scenario in ConCap.,
label={lst:replicated-detgen},
float=t
]
attacker:
  name: siege
  image: ghcr.io/idlab-discover/concap/siege:ubuntu18
  atkCommand: siege -c 10 -r 1000 -v http://$TARGET_IP
  atkTime: 10s
target:
  name: nginx
  image: nginx:1.13.8-alpine
\end{lstlisting}


\begin{lstlisting}[
frame=single,
breaklines=true, 
basicstyle=\ttfamily\footnotesize,
caption=A minimal container image for running \textit{siege}.,
label={lst:siege-image},
float=t
]
FROM ubuntu:18.04
ENV DEBIAN_FRONTEND noninteractive

RUN apt-get update && \
    apt-get -y install siege && \
    apt-get clean && \
    rm -rf /var/lib/apt/lists 

ENTRYPOINT ["siege"]
\end{lstlisting}

\begin{lstlisting}[
frame=single,
breaklines=true, 
basicstyle=\ttfamily\footnotesize,
caption=A \concap{} multi-target scenario file defining a multi-step attack chain against three targets. The scenario combines global with target-specific network and labeling configuration.,
label={lst:multi-target-scenario},
basicstyle=\footnotesize,
float=t
]
type: multi-target
name: multi-step-attack
attacker:
  name: advanced-attacker
  image: ghcr.io/idlab-discover/concap/advanced-attacker:1.0.0
  atkCommand: ./multi-step-attack.sh $TARGET_IP_0 $TARGET_IP_1 $TARGET_IP_2
  cpuRequest: 100m
  memRequest: 250Mi
targets:
  - name: openssh
    image: ghcr.io/idlab-discover/concap/openssh-server:password-24.04
    cpuRequest: 200m
    memRequest: 200Mi
    labels:
      step: "bruteforce-ssh"
    network:
      bandwidth: 10Mbit
      queueSize: 100ms
      delay: 100ms
  - name: db
    image: ghcr.io/idlab-discover/concap/mysql:1.0.0
    cpuRequest: 200m
    memRequest: 200Mi
    labels:
      step: "exfiltration"
    network:
      bandwidth: 1Gbit
      queueSize: 100ms
      delay: 1ms
  - name: libssh
    image: vulhub/libssh:0.8.1
    cpuRequest: 200m
    memRequest: 200Mi
    labels:
      step: "exploit-cve"
network:
  bandwidth: 100Mbit
  queueSize: 100ms
  delay: 5ms
labels:
  label: 1
  category: "advanced-lateral"
  
\end{lstlisting}

\subsection{Configuration of a SOCBED Scenario in \concap{\large}}
\label{sapp:socbed}
\noindent
We further show the coverage of \concap{} by providing the configuration files that enable reproduction of the experimental setup of SOCBED~\cite{uetz2021reproducible}. 

To this end, we report in Listing~\ref{lst:socbed} the configuration that reproduces a portscan, i.e., the first step of the attack chain in SOCBED. We also provide the corresponding PCAP and (labeled) NetFlow data in our repository (note: this data is \textit{not} provided by SOCBED).
Furthermore, in our repository~\cite{repository}, we report additional configuration files (and corresponding data) for other scenarios in SOCBED. 

We observe that creating all these configuration files required us less than one day of work: we merely investigated the public repository of SOCBED~\cite{socbedGithub} to infer the low-level network and command details, required to devise our own configuration files.

\begin{lstlisting}[
frame=single,
breaklines=true, 
basicstyle=\ttfamily\footnotesize,
caption=A configuration of \concap{} to support one of the scenarios envisioned in SOCBED which includes a portscan.,
label={lst:socbed},
basicstyle=\footnotesize,
float=t
]
type: multi-target
name: socbed-example
attacker:
  name: attacker
  image: instrumentisto/nmap
  atkCommand: nmap $TARGET_IP_1 $TARGET_IP_2 $TARGET_IP_3 $TARGET_IP_4 $TARGET_IP_5 -n --disable-arp-ping -sU -sV
  cpuRequest: 200m
  memRequest: 200Mi
targets:
  - name: dmz-server
    image: vulnerables/web-dvwa
    cpuRequest: 200m
    memRequest: 200Mi
    labels:
      server: "dmz-server"
  - name: log-server
    image: kibana:9.2.1
    cpuRequest: 200m
    memRequest: 200Mi
    labels:
      server: "log-server"
  - name: internal-server
    image: dperson/samba
    cpuRequest: 200m
    memRequest: 200Mi
    labels:
      server: "internal-server"
  - name: client-1
    image: dockurr/windows
    cpuRequest: 200m
    memRequest: 200Mi
    labels:
      server: "client-1"
  - name: client-2
    image: dockurr/windows
    cpuRequest: 200m
    memRequest: 200Mi
    labels:
      server: "client-2"
network:
  bandwidth: 100Mbit
  queueSize: 100ms
  delay: 5ms
labels:
  label: 1
  category: "socbed-example"

  
\end{lstlisting}

\subsection{Scenario Details for Broad Analysis \concap{\Large}}
\label{sapp:commands-concap}
\noindent
Each scenario evaluates a specific network activity or tool using its default configuration unless noted otherwise. Where applicable, we highlight the use of additional command options. All scenario files, including execution and configuration details, are available in our repository for full reproducibility.

\textbf{Wfuzz} This scenario simulates an HTTP fuzzing attack where the attacker attempts to discover hidden web resources by injecting path variations. The ''common'' wfuzz wordlist was used as an additional parameter.

\textbf{Slowloris} This activity performs a Slowloris denial-of-service attack, in which the attacker opens many half-completed HTTP connections to exhaust the server’s resources. The attack was configured to run for 300 seconds, beyond the tool's default behavior.

\textbf{Patator FTP} This scenario emulates an FTP brute-force attack using Patator. A password list with the 200 most used passwords of 2023~\cite{danielmiesslerseclists} was explicitly supplied, and failure responses matching the message "530 Login incorrect." were ignored using Patator’s ignore option.

\textbf{Ping} A standard ICMP echo test is performed with 10 packets sent to the target host. The number of packets was explicitly specified in the command.

\textbf{Patator SSH} This scenario represents a brute-force attack against an SSH server. Both a username and password list were provided, top usernames shortlist and top 200 passwords from 2023~\cite{danielmiesslerseclists}, and the failed login attempts are filtered based on the message "Authentication failed."

\textbf{Nmap Version Scan} In this scan, Nmap was used with additional options for version detection, SYN scan, and no DNS resolution. The scan targeted ports 79 and 80.

\textbf{Nmap} This scenario runs a more basic Nmap scan, using a SYN scan and disabling DNS resolution. It targeted the same ports (79 and 80) but did not include version detection.

\textbf{MySQL} The attacker connects to a MySQL database and issues a query to read entries from a users table. Default MySQL authentication is used (user: root, password: root), and the server was configured to listen on all interfaces.

\textbf{Curl/FTP} This scenario involves an authenticated file download over FTP. The attacker connects with a predefined username and password to retrieve a specific file (bible.txt). No additional options were used beyond basic authentication and output specification.

\textbf{Dig (DNS)} The attacker sends a series of DNS queries for five popular domain names (e.g., google.com, facebook.com) to the target DNS server on a non-standard port (63). The use of a custom port deviates from the default behavior.
\section{Datasets (existing and new ones)}
\label{app:data}

\noindent
We provide details on the various ``datasets'' considered in our paper. Specifically, we first provide information on the real-world capture (used in §\ref{ssec:development}), then we describe how we preprocessed the benchmark datasets CICIDS17 and CICIDS18 (used in §\ref{ssec:replicability} and §\ref{ssec:enhancing}), and finally we provide low-level details of our ``new'' dataset containing labeled NetFlows related to attacks not contained in currently available benchmarks (mentioned in §\ref{ssec:enhancing}).

\subsection{Real-word Network Capture: Description}
\label{sapp:homenetwork}

\noindent
For the experiments in §\ref{ssec:development}, we used data captured in a real-world network representing a ``smart home''. Here, we provide more details of this network environment, and the captured PCAP trace and corresponding NetFlow data.

\textbf{Network Overview}
The network environment encompasses 40--50 physical devices. Such devices entail: smartphones, laptops / desktops, gaming consoles; as well as various IoT devices (smart speakers, lightbulbs) and media appliances (e.g., smart TV). All these devices are connected to a router through a WiFi 5 or 2.4 interface. The router is connected to the internet through a 50Mbps download speed and 5Mbps upload speed. The devices within the network are kept up-to-date with security patches and their owners are security experts, hence it is safe to assume that the traffic is ``benign'' (and even if some traffic is ``malicious'', it is of a different class than \command{patator}{\small} meaning that our conclusions are not affected by such circumstances). With regards to the devices used to simulate the attack in this network, they were two laptops both running Ubuntu 18.04; the ``target'' mounts an Intel i7 7700H with 32GB of RAM; the ``defender'' mounts an Intel N4100 with 8GB of RAM.

\textbf{Network Traffic (PCAP).}
Three sets of network captures were performed, two for benign background traffic and one related to the \command{patator}{\small} ssh-bruteforce attacks. The background traffic was first captured on the 6th of November 2023 and a second time on the 26th of August 2024, the same day the malicious attacks were executed and captured. The total file size for the benign traffic of the August 26th capture is of 6GB for 8M packets, whereas the attack captures only measured 45MB for 241k packets. For the (benign) trace captured on November 2023, the size is of 17GB for 17M packets

\textbf{NetFlow.}
NetFlows were extracted from the network captures using CICFlowMeter. The benign capture from November 2023 contained 30,304 unique NetFlows, while the benign capture from August 2024 contained 49,188 unique NetFlows. The malicious capture, on the other hand, included a total of 7,990 unique NetFlows.

\subsection{Preprocessing of Benchmark datasets}
\label{sapp:benchmark}

\noindent
In this appendix, we provide an overview of the data preprocessing steps undertaken to prepare the network traffic data for analysis. This includes the extraction of NetFlow features using the latest version of CICFlowMeter, data cleaning procedures, and a description of the real world network captures used in our experiments.

\textbf{NetFlow Extraction}
The fixed versions of the datasets CICIDS17 and CICIDS18 were released in October 2022. Since then, CICFlowMeter, the tool used to extract the NetFlow features from the network traffic traces, has received over 30 new commits fixing, changing, and adding NetFlow features. To benefit from these updates over the last two years, we replicated the work done by Liu et al.~\cite{liu2022error} by extracting the NetFlows using the current version (Aug '24) of CICFlowMeter and labeling the flows based on their fixed logic. The ``attempted'' NetFlows are removed from our dataset.

\textbf{Data Cleaning.}
The dataset cleaning steps on the NetFlows performed in our experiments follow the best practices described in previous work~\cite{apruzzese2022sok, arp2022and, d2022establishing}. First, the meta-data and spurious features are removed: ``id'', ``Flow ID'', ``Src IP'', ``Dst IP'', ``Timestamp'', ``FWD Init Win Bytes'', and ``Bwd Init Win Bytes''. Second, the source and destination ports are mapped to their IANA port categories and subsequently encoded as 0, 1, or 2 for respectively \textit{well-known}, \textit{registered}, and \textit{dynamic}. Third, all NetFlows with missing values are removed. Last, all duplicate NetFlows are removed. 

\textbf{Implementation.}
The features and hyperparameters used to develop our ML models are provided in our repository.

\subsection{Generation of (new) CVE Data with \concap{\large}}
\label{sapp:new_data}

\noindent
We generated new labeled "malicious" data using \concap{} by executing three recent CVE-based attack scenarios, ensuring none of them overlap with the datasets mentioned in Flood et al.~\cite{flood2024bad}. These CVEs represent diverse vulnerabilities, each exploited to demonstrate the flexibility of \concap{} in handling modern attacks.

\begin{itemize}[leftmargin=*]
    \item \textbf{CVE-2024-47177}~\cite{cve1}: This vulnerability affects OpenPrinting Cups-Browsed versions 2.0.1 and earlier, where improper handling of the FoomaticRIPCommandLine parameter in PostScript Printer Description (PPD) files allows remote code execution. Attackers can exploit this by creating a malicious IPP (Internet Printing Protocol) server that sends crafted printer information to a vulnerable Cups-Browsed instance, enabling arbitrary command execution on the affected system. 
    \item \textbf{CVE-2024-36401}~\cite{cve2}: In GeoServer versions prior to 2.25.1, 2.24.3, and 2.23.5, unauthenticated remote code execution is possible through unsafely evaluated property name expressions. Specially crafted input can exploit these unsafe evaluations in multiple OGC request parameters, allowing attackers to execute arbitrary code on a vulnerable GeoServer installation. 
    \item \textbf{CVE-2024-2961}~\cite{cve3}: This vulnerability in the GNU C Library (versions 2.39 and earlier) affects the iconv() function, which may overflow the output buffer by up to 4 bytes when converting strings to the ISO-2022-CN-EXT character set. Attackers can exploit arbitrary file read vulnerabilities in PHP applications to escalate to remote code execution by leveraging the iconv() issue, potentially crashing the application or executing code.
\end{itemize}

The target environments were constructed using Vulhub, an open-source collection of pre-built vulnerable docker environments. The attacker environments were built using official Docker images, combined with the necessary tools and exploit code for each CVE. These environments were then defined into a \concap{} scenario together with the attack command and assigned a unique label with the corresponding CVE for automatic traffic labeling. The three scenarios, one for each CVE, are then repeated 320 times in different network environments with a unique set of values for \textit{delay, loss, corrupt,} and \textit{duplicate}. Each of these scenarios is repeated for five, two, and four host-space perturbations respectively for CVE-2024-47177, CVE-2024-36401, and CVE-2024-2961.

\begin{table}[!htbp]
    \centering
    \caption{\textbf{Testing existing ML-NIDS against unseen CVEs with \concap{\footnotesize}.} \textmd{We develop benchmark ML-NIDS using CICIDS17 (baseline \scmath{tpr}=0.999 with \scmath{fpr}=0.001) and CICIDS18 (baseline \scmath{tpr}=0.999 with \scmath{fpr}=0.001). Then, we use \concap{\footnotesize} to generate and label NetFlows of three CVEs (we report the \# Packets and \# NetFlows of each capture) involving completely different attacks than those used to ``train'' the ML-NIDS, and we test them against all of our models (RF, DT, SVM, DNN, HGB). The results show the average \scmath{tpr} (and std. dev.) across all models.}}
    \label{tab:new_data}
    \vspace{-2mm}
    \resizebox{0.95\columnwidth}{!}{
        \begin{tabular}{c?c|c?c|c}
            \toprule
            & \multicolumn{2}{c?}{Traffic Statistics} & CICIDS17 & CICIDS18 \\
            \textbf{Attack} & \# \textbf{Packets} & \# \textbf{NetFlows} & \smamath{tpr} (std. dev) & \smamath{tpr} (std. dev) \\
            \midrule
            CVE-2024-47177 & 44658 & 4800 & 0.205 (0.389) & 0.550 (0.396)\\
            CVE-2024-36401 & 10372 & 640 & 0.199 (0.399) & 0.214 (0.378)\\
            CVE-2024-2961 & 391350 & 1280 & 0.112 (0.224) & 0.011 (0.021)\\
            
            \bottomrule
        \end{tabular}
    }
    \vspace{-3mm}
\end{table}

\section{Additional Results and Experiments}
\label{app:results}

\noindent
We presents detailed results for the interested reader.

First, we provide the training time of the models evaluated on the real-world network in Table~\ref{tab:demo_real_timing}. Finally, Fig.~\ref{fig:feature_dist} shows an additional 30 NetFlow features compared for the patator brute force attack between traffic generated by \concap{} and on a real network as described in §\ref{sec:validation}. 

We provide the tables with the standard deviation of our results (useful to carry out statistical tests) in our repository~\cite{repository}.

\subsection{Runtime and Computing Results}
\label{sapp:runtime}

\noindent
We report in Table~\ref{tab:demo_real_timing} the training time of the ML models used for the experiments in §\ref{sec:application}. These experiments have been carried out on a machine running Windows 11 Pro (Build 26100) with a Intel 12th Gen Core i7-1265U CPU \@ 1.80 GHz (10 cores, 12 threads) and 32GB RAM.

Then, we report in Table~\ref{table:concap-perf} the results of the runtime assessment of \concap{} (covered in §\ref{ssec:runtime}). The laptop used in these tests runs MicroK8s v1.32.3 in a virtualized Linux environment on Windows Subsystem for Linux 2 (WSL2). The host system features an Intel 12th Gen Core i7-1265U CPU \@ 1.80 GHz (10 cores, 12 threads), 32GB RAM, and Windows 11 Pro (Build 26100), WSL2 has access to 12 logical CPUs and 23 GiB of RAM.
During our tests, 12 variations of \command{nmap}{\small} scenarios are executed in a single \concap{} run configured to run three scenarios concurrently. 
\concap{} shows minimal differences in initialization time and resource usage across the two setups. Interestingly, initialization time is slightly lower on the laptop despite its more limited resources. We attribute this to the elimination of network communication delays between the machine running \concap{} and the cluster executing the scenarios, as the microK8s setup is entirely local. Initialization time includes spawning the attacker and target containers, configuring the scenario network environment, and activating packet capture on the targets, as well as associated network delays. We also measured the resource consumption of the processing pods: both \command{argus}{\small} and \command{cicflowmeter}{\small}. While \command{argus}{\small} remains lightweight, \command{cicflowmeter}{\small} consistently exhibits an order of magnitude higher CPU and memory usage compared to the scanning scenarios themselves---highlighting the relative cost of traffic processing versus generation. Although both environments perform similarly, the Kubernetes cluster provides horizontal scalability, enabling the execution of an increased number of scenarios in parallel when needed.

\begin{table}[h]
    \caption{\textbf{Training times for the models for the assessment of the real-world network.} \textmd{\footnotesize Avg time (sec) and std. dev. over 5 training folds.}}
    
    \label{tab:demo_real_timing}
    \vspace{-1mm}
    \begin{subtable}{\columnwidth}
    \centering
    \resizebox{\columnwidth}{!}{
        \begin{tabular}{
        l |
        S[table-format=2.1(1.1)] S[table-format=2.1(1.1)] 
        }
        \toprule
         
        Train Set & \multicolumn{1}{c}{Benign + P=1} & \multicolumn{1}{c}{Benign + P=0} \\ 
        \midrule
        DT & 0.3(0.0) & 0.2(0.0) \\ 
        RF & 3.1(0.1) & 4.1(0.2) \\ 
        XGB & 3.0(0.4) & 3.9(0.6) \\ 
        SVM & 0.5(0.0) & 0.9(0.1) \\ 
        DNN & 9.2(0.5) & 12.3(0.5) \\ 

        \bottomrule
        \end{tabular}
    }
    \caption{Real-world}
    \label{tab:timing-real}
    \end{subtable}
    \hfill
    \begin{subtable}{.485\columnwidth}
    \centering
    \resizebox{\columnwidth}{!}{
        \begin{tabular}{
        l |
        S[table-format=2.1(1.1)] 
        }
        \toprule
         & \multicolumn{1}{c}{\textbf{Train time}} \\ 
        \midrule
        DT & 6.7(1.3) \\ 
        RF & 40.8(4.2) \\ 
        XGB & 15.1(3.8) \\ 
        SVM & 4.1(0.9) \\ 
        DNN & 17.2(2.0) \\ 

        \bottomrule
        \end{tabular}
    }
    \caption{CICIDS17}
    \label{tab:timing-cic17}
    \end{subtable}
    \hspace{\fill}
    \begin{subtable}{.485\columnwidth}
    \centering
    \resizebox{\columnwidth}{!}{
        \begin{tabular}{
        l |
        S[table-format=3.1(2.1)] 
        }
        \toprule
         & \multicolumn{1}{c}{\textbf{Train time}} \\ 
        \midrule
        DT & 40.4(1.8) \\ 
        RF & 547.8(6.3) \\ 
        XGB & 118.8(7.4) \\ 
        SVM & 101.3(10.9) \\ 
        DNN & 574.5(45.8) \\ 

        \bottomrule
        \end{tabular}
    }
    \caption{CICIDS18}
    \label{tab:timing-cic18}
    \end{subtable}
    \vspace{-2mm}
\end{table}

\begin{table}[ht]
\centering
\caption{\textbf{Performance evaluation of \concap{} running various port scan scenarios on a multi-node K8s cluster versus microK8s on a WSL-enabled laptop.}
The results show minimal differences in initialization time and resource usage between the two environments. However, the K8s cluster offers horizontal scalability, allowing for greater resource availability and enabling more parallel scenario executions.}
\label{table:concap-perf}
\vspace{-2mm}
\resizebox{\columnwidth}{!}{
\begin{tabular}{l | r r r | r r r}
\toprule
 & 
\multicolumn{3}{c|}{\textbf{Cluster}} & 
\multicolumn{3}{c}{\textbf{Laptop}} \\
Scenario & Init Time (s) & CPU (mcpu) & Mem (MiB) & Init Time (s) & CPU (mcpu) & Mem (MiB) \\
\midrule
nmap-pn-ss       & 2.202 & 4.10  & 1.67 & 2.300 & 2.77  & 0.68 \\
nmap-pn-ss-sv    & 2.696 & 20.94 & 1.67 & 2.155 & 2.67  & 2.00 \\
nmap-pn-st       & 2.622 & 7.89  & 1.67 & 1.866 & 4.37  & 1.63 \\
nmap-pn-st-sv    & 2.737 & 3.33  & 1.67 & 2.189 & 2.53  & 2.00 \\
nmap-pn-su       & 2.747 & 3.33  & 1.67 & 2.095 & 4.52  & 1.67 \\
nmap-pn-su-sv    & 2.724 & 19.76 & 1.67 & 2.869 & 1.00  & 1.67 \\
nmap-ss          & 2.679 & 3.00  & 1.67 & 2.300 & 1.00  & 1.67 \\
nmap-ss-sv       & 4.682 & 10.88 & 1.67 & 2.155 & 11.03 & 2.00 \\
nmap-st          & 2.286 & 4.67  & 1.67 & 1.866 & 4.40  & 1.67 \\
nmap-st-sv       & 3.996 & 15.17 & 1.67 & 2.189 & 8.71  & 2.00 \\
nmap-su          & 3.767 & 8.00  & 1.67 & 2.095 & 3.00  & 1.67 \\
nmap-su-sv       & 2.536 & 1.00  & 1.67 & 2.869 & 8.00  & 1.67 \\
\midrule
argus            & ---   & 18.19 & 0.00 & ---   & 14.88 & 0.00 \\
cicflowmeter     & ---   & 119.85& 2.98 & ---   & 109.10& 10.95 \\
\midrule
\concap{}        & ---   & 33.18 & 35.29& ---   & 29.95 & 33.76 \\
\midrule
\textbf{Avg Init Time} & \multicolumn{3}{c|}{2.97 s} & \multicolumn{3}{c}{2.25 s} \\
\bottomrule
\end{tabular}
}
\end{table}

\begin{figure}[tbp]
    \centering
    \includegraphics[width=\columnwidth]{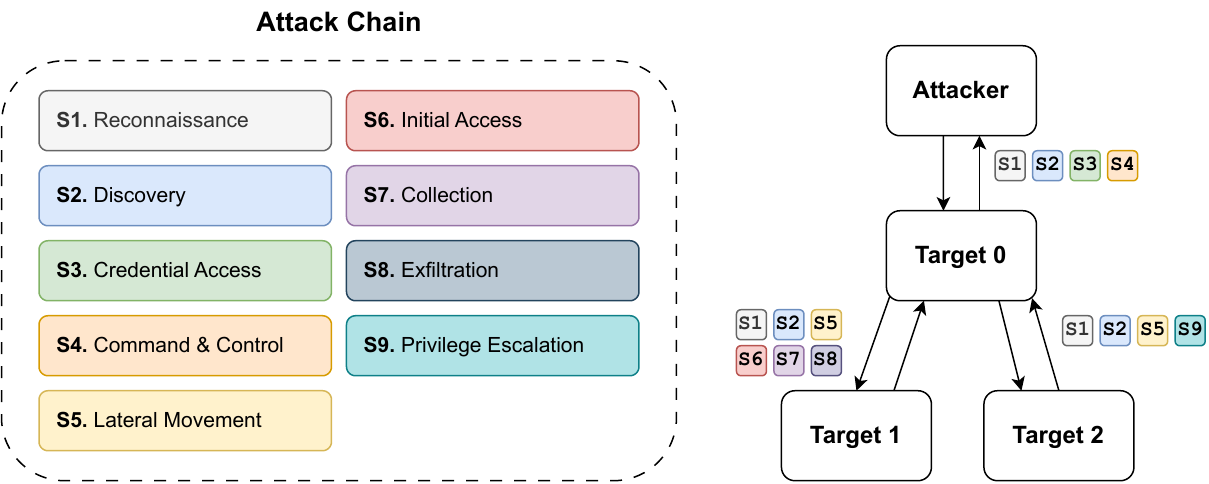}
    
    \caption{{\textbf{Using \concap{} to reproduce complex attack chains envisioned in MITRE ATT\&CK.}
    \textmd{{\footnotesize The attacker first compromises an exposed target before performing lateral movement to two internal hosts.}}}} 
    \label{fig:multi}
    \vspace{-6mm}
\end{figure}

\subsection{Reproducing Multi-step and Multi-host attacks}
\label{sapp:multistep}
\noindent
We conclude our practical demonstrations by showing how to use \concap{} to carry out experiments entailing sophisticated attacks. Specifically, we demonstrate that \concap{} enables automatic reproduction of multi-host and multi-step attacks. 

\textbf{Threat Model.} We assume a network consisting of four or more hosts, each having different privileges and containing different information. An attacker has obtained control of one host, and wants to expand their control and steal sensitive information. However, the attacker does not know the network topology, and must hence carry out reconnaissance activities to identify vulnerabilities that can be exploited to reach the intended goal. In other words, we envision an attack chain spanning multiple stages and hosts, which can be mapped to a range MITRE ATT\&CK tactics~\cite{mitreattack}. Altogether, the attack encompasses the following stages (expressed via the well-known MITRE ATT\&CK terminology): Gather Victim Network Information, Network Service Discovery, Brute Force, Internal Proxy, SSH Lateral Movement, Data Collection from Local System, Exfiltration Over C2, and Exploitation for Privilege Escalation. Our intention is showing how to use \concap{} to generate (and automatically label) NetFlows of the entire attack chain---in an end-to-end fashion.

\textbf{Implementation.} To replicate the setting envisioned in our threat model, we defined the scenario in Listing~\ref{lst:multi-target-scenario} (in the Appendix). Such a scenario embeds the use case, schematically depicted in Fig.~\ref{fig:multi}, in which an attacker first performs a port scan against an SSH server using \command{nmap}{\small}~\cite{nmap}, then launches a brute-force SSH attack using \command{Medusa}{\small}~\cite{medusa}. Upon discovering valid credentials, the attacker uses SSH to set up a SOCKS proxy through the compromised host and scans additional internal hosts (a MySQL database and a vulnerable libssh server). Then, the attacker forwards internal ports via SSH tunneling to the local machine, allowing authenticated access and data exfiltration against the MySQL target and enabling the attacker to exploit a known vulnerability against libssh.

\textbf{Considerations.}
In practice, when \concap{} processes such a scenario, all of the aforementioned operations are automatically carried out. 
We also note that each target in this scenario is separately labeled with its function (e.g., ``brute-force-ssh'', ``exfiltration'', or ``cve'') and configured with specific network constraints (e.g., bandwidth and delay). At the end of the entire experiment, the OpenSSH server received 3635 network packets corresponding to 184 CICFlowMeter and 185 argus NetFlows. The MySQL server received 280 packets corresponding to 109 CICFlowMeter and 110 argus NetFlows. The libSSH server received 246 packets corresponding to 106 CICFlowMeter and 106 argus NetFlows. All NetFlows (which we provide in our repository~\cite{repository}) are labeled according to the specifications of the scenario. Notably, by sharing this scenario (i.e., Listing~\ref{lst:multi-target-scenario}), future researchers can inspect the entire traffic-generation process: this is important for open science and to address the skepticism around NIDS research~\cite{flood2024bad}.

\subsection{Testing additional ML-based NIDS with \concap{}}
\label{sapp:contemporary}
\noindent
In our paper (§\ref{sec:application}), we showed that \concap{} can be used to test ``simple'' ML-based (e.g., RF or SVM) NIDS. Here, we expand our assessment by showing that \concap{}'s generated data can be used also by other families of ML-driven NIDS.

\uline{Please note} that our goal is merely to show that \concap{}'s data can be used to train and test such additional methods: we do not seek to {\small \textit{(a)}}~propose new methods, {\small \textit{(b)}}~outperform/tweak existing ones, or {\small \textit{(c)}}~benchmark prior work.

\textbf{Considered models.} There are hundreds of papers proposing, or evaluating, methods to detect network intrusions via ML~\cite{apruzzese2023sok}. For the sake of this additional demonstration, we consider the methods considered in the recent USENIX'23 paper xNIDS~\cite{wei2023xnids}. Specifically, this work considers two families of deep-learning--based methods for intrusion detection: one reliant on \textit{AutoEncoders} (drawn from~\cite{mirsky2018kitsune,shone2018deep}), which we denote as AE-IDS; and another one reliant on \textit{Recurrent Neural Networks} (drawn from~\cite{yin2017deep,jan2020throwing}), which we denote as RNN-IDS. Note that, altogether, the papers proposing these methods have thousands of citations on Google Scholar (as of December 2025) or are published in top-tier venues (e.g., S\&P'22, NDSS'18). Then, to provide yet-another perspective, we also consider transformer-based approaches which empower \textit{multimodal large-language models} (LLMs), inspired by the recent~\cite{luay2025multimodal}. 

\textbf{Implementation and Datasets.} We take the code from~\cite{wei2023xnids} (which is publicly available) as a blueprint for our AE-IDS and RNN-IDS models; whereas, for LLMs, we consider: ChatGPT~5.1, and Gemini~Flash~2.5 (both of which being the best freely-available commercial models as of December 2025). For both of these LLMs, inspired by~\cite{luay2025multimodal}, we will assess them with a zero-shot prompting strategy (with the prompt in Listing~\ref{lst:chatgpt_zero}), as well as in an ``zero-shot-with-augmentation'' fashion (with the prompt in Listing~\ref{lst:chatgpt_aug}).\footnote{Note that the approach in~\cite{luay2025multimodal} uses \textit{two-dimensional plots} as input to the LLM, whereas we use \concap{}'s raw data. This is because we want to show that \concap{} can be used off-the-shelf to test LLMs. However, there is nothing preventing one from generating the same two-dimensional plots of~\cite{luay2025multimodal} (which simply show the source and destination IP for the x- and y-axes, and points indicate the amount of \textit{exchanged bytes}---all of which being metrics derivable from \concap{}'s data.} For the evaluation dataset, we consider the same used in §\ref{ssec:replicability}, i.e., a mix of CICIDS17 and the data generated by \concap{} referring to the ssh-\command{patator}{\small} attack. Such a setup is valid for the sake of our demonstration, given that the approaches we considered have been evaluated in completely different setups (e.g., the ``outdated'' NSL-KDD or the CIC-DoS2017~\cite{jazi2017detecting}). 

\textbf{Evaluation and Results.} We replicate the experiments in §\ref{ssec:replicability}. Specifically, we train each model (AE-IDS and RNN-IDS) on 80\% of the benign data in CICIDS17, as well as on 80\% of the data generated by \concap{} for ssh-\command{patator}{\small} (launched with default options); and then we test it on the remaining 20\% (benign and malicious). We also consider doing a mixed experiment by testing each model on 20\% of the ssh-\command{patator}{\small} included in CICIDS17; as well as by creating another variant trained on 80\% of the ssh-\command{patator}{\small} included in CICIDS17 and testing it on 20\% of the ssh-\command{patator}{\small} generated by \concap{}. For the LLMs, the zero-shot prompt does not have any training (barring that to develop the LLM itself), whereas the augmented-zero-shot prompt embeds a fine-tuning step done on the same data used to train the AE-IDS and RNN-IDS. We repeat our experiments five times to ensure consistency. Let us discuss the results of our experiments:
\begin{itemize}[leftmargin=*]
    \item \textit{AE-IDS:} This experiment confirms our conclusions drawn from §\ref{ssec:replicability}. Specifically, when trained\footnote{Training the AE-IDS required $\approx$10 minutes on our system.} on \concap{}'s generated data (which is always malicious), the AE-IDS achieves \smamath{tpr}=0.999 on the testing partition of \concap{}'s generated data, and \smamath{tpr}=0.987 on the testing partition of malicious data of the same attack included in CICIDS17; conversely, when trained on the malicious data of CICIDS17, the AE-IDS achieves \smamath{tpr}=0.955 on the testing partition of CICIDS17, and \smamath{tpr}=0.952 on the testing partition of \concap{}'s generated data. The \smamath{fpr} is 0.478 on the former case, and 0.209 in the latter case: such an underwhelming result is because our AE-IDS uses the same thresholding mechanism used in xNIDS~\cite{wei2023xnids} (note that AutoEncoders are not classifiers: to use them in a classification task, one must specify a threshold on the reconstruction error. We used the one of xNIDS, which was derived on a different dataset. To improve the \smamath{fpr}, one can simply change the threshold). The different \smamath{fpr} is because the number of datapoints in the training set is different across the two experiments, but what is crucial is that, for this detector, \uline{the data generated by \concap{} is essentially equivalent to that of CICIDS17}, since the classification results on these two groups are statistically the same (verified with a t-test: \smamath{p<.05}).

    \item \textit{RNN-IDS:} For this experiment, we report that, after 2 hours of training, the underlying RNN did not reach convergence. Therefore, given our previous experiment, as well as those in our main paper (§\ref{ssec:replicability}) showing that simple models (such as RF or SVM) can achieve near-perfect detection accuracy while requiring seconds to train (see Table~\ref{tab:timing-cic17}), we conclude that using RNN-IDS on our dataset is not a wise choice and stopped this experiment. However, what matters is that the RNN-IDS was being trained, meaning that \concap{}'s data was ``compatible'' with that expected in RNN-IDS, thereby validating the goal of these experiments (note that SaTML'26 call for papers explicitly states to be mindful of the computing resources spent in training ML models, and we believe that additional training would not have changed our conclusions).

    \item \textit{ChatGPT:} When assessed in the ``zero-shot'' prompt, the model always classified each datapoint as benign. This is expected: the malicious datapoints are just repeated SSH attempts, which not necessarily imply malicious intent. However, by using the ``augmented-zero-shot'' prompt (which uses \concap{}'s generated labelled data to ``teach'' the LLM that quick bursts of SSH connections are malicious), the model achieves a perfect classification accuracy (i.e., \smamath{tpr}=1.000 and \smamath{fpr}=0.000). We inspected the output provided by ChatGPT, and found that, to derive such a result, the model learned heuristics that enabled a clear separation of the benign from malicious datapoints. (note that these results align with those in~\cite{luay2025multimodal}).

    \item \textit{Gemini:} The results for Gemini 2.5 match those of ChatGPT.
\end{itemize}
The code of these experiments is in our repository~\cite{repository}.

\begin{cooltextbox}
\textsc{\textbf{Takeaway.}} Our experiments show that \concap{} can be used to develop and assess multiple families of ML-based NIDS. Moreover, our results further confirm that \concap{}'s generated data can be functionally equivalent to that generated in other environments (e.g., that of CICIDS17).
\end{cooltextbox}

\begin{lstlisting}[
frame=single,
breaklines=true, 
basicstyle=\ttfamily\footnotesize,
caption=Zero-shot prompt used to evaluate LLMs with \concap{}'s data.,
label={lst:chatgpt_zero},
basicstyle=\footnotesize,
float=t
]
You are a cybersecurity analyst working in a Security Operations Center (SOC). You are analyzing a CSV file that contains NetFlow records. 

Data Description: 
Each row in the CSV represents a network flow with features such as: Src Port, Dst Port, Protocol, Flow Duration, Number of packets Your task is to classify each row as either: 0 = Benign 1 = Malicious 

Your Tasks: 
Determine whether the NetFlow record shows suspicious behavior. Output only 0 or 1 for each row (0 = benign, 1 = malicious) on a separate row and write to csv file.
\end{lstlisting}

\begin{lstlisting}[
frame=single,
breaklines=true, 
basicstyle=\ttfamily\footnotesize,
caption=Augmented zero-shot prompt used to evaluate LLMs with \concap{}'s data. We augment the prompt in Listing~\ref{lst:chatgpt_zero} by instructing the LLM to use labeled data (in the X\_train.csv file) as a guide for its inference.,
label={lst:chatgpt_aug},
basicstyle=\footnotesize,
float=t
]
You are a cybersecurity analyst working in a Security Operations Center (SOC). You are analyzing a CSV file that contains NetFlow records (X_test_shuffled.csv). 

Data Description:
Each row in the CSV represents a network flow with features such as: Src Port, Dst Port, Protocol, Flow Duration, Number of packets Your task is to classify each row as either: 0 = Benign 1 = Malicious 

Your Tasks: 
Determine whether the NetFlow record shows suspicious behavior. Base your predictions on the provided labeled NetFlows (X_train.csv and y_train.csv). Output only 0 or 1 for each row (0 = benign, 1 = malicious) on a separate row and write to csv file.
\end{lstlisting}

\begin{figure*}[t]\captionsetup[subfigure]{font=footnotesize,skip=0pt}
    \centering
    \begin{subfigure}[b]{0.18\textwidth}
        \centering
        \includegraphics[width=\textwidth]{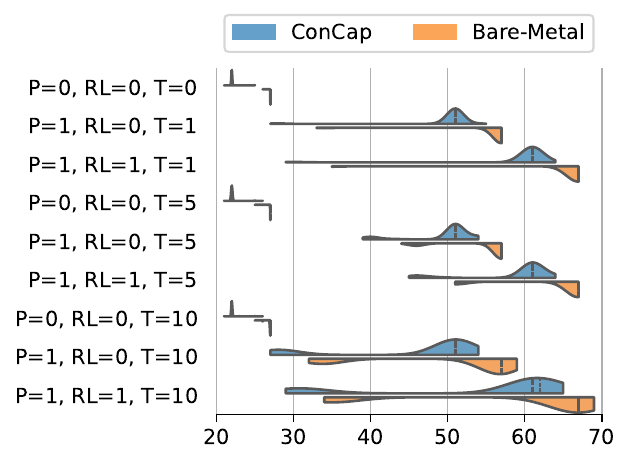}
        \caption*{Ack Flag Count}
    \end{subfigure}
    \begin{subfigure}[b]{0.18\textwidth}
        \centering
        \includegraphics[width=\textwidth]{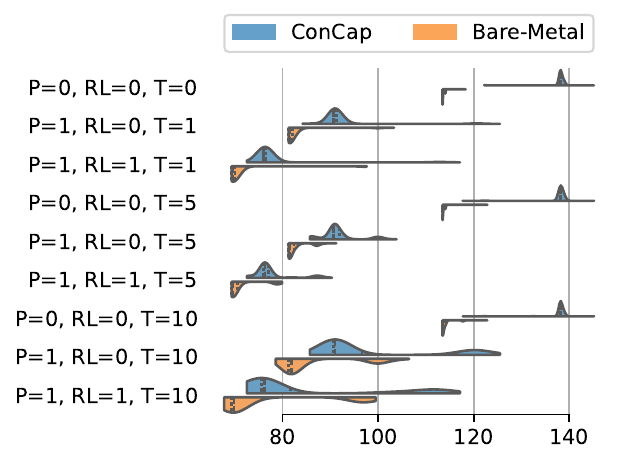}
        \caption*{Average Packet Size}
    \end{subfigure}
    \begin{subfigure}[b]{0.18\textwidth}
        \centering
        \includegraphics[width=\textwidth]{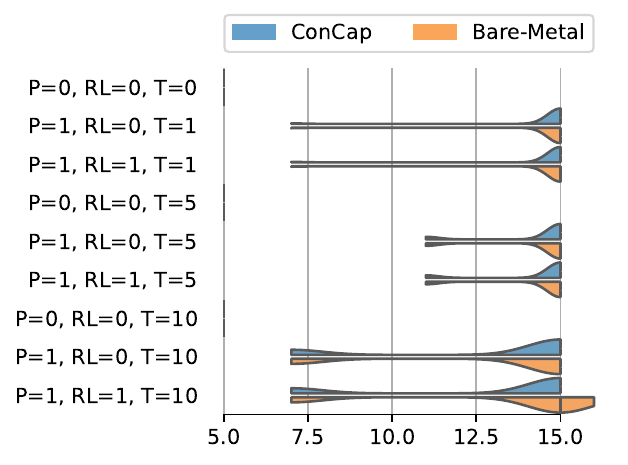}
        \caption*{Bwd Act Data Packets}
    \end{subfigure}
    \begin{subfigure}[b]{0.18\textwidth}
        \centering
        \includegraphics[width=\textwidth]{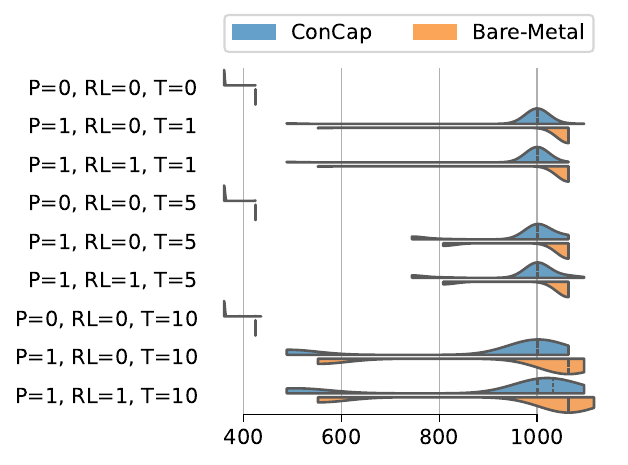}
        \caption*{Bwd Header Length}
    \end{subfigure}
    \begin{subfigure}[b]{0.18\textwidth}
        \centering
        \includegraphics[width=\textwidth]{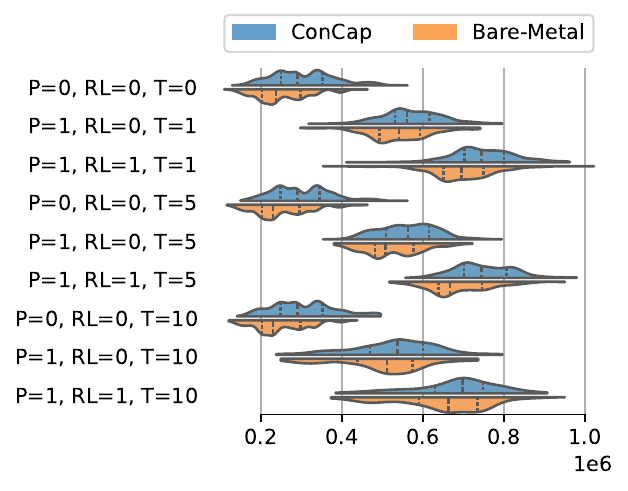}
        \caption*{Bwd IAT Mean}
    \end{subfigure}
    \par\medskip
    \begin{subfigure}[b]{0.18\textwidth}
        \centering
        \includegraphics[width=\textwidth]{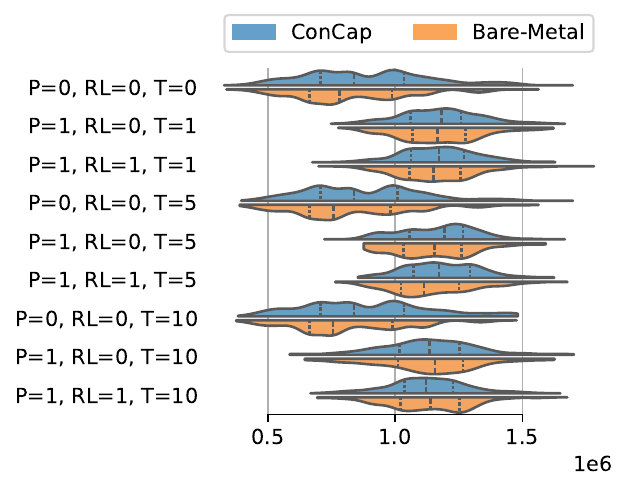}
        \caption*{Bwd IAT Std}
    \end{subfigure}
    \begin{subfigure}[b]{0.18\textwidth}
        \centering
        \includegraphics[width=\textwidth]{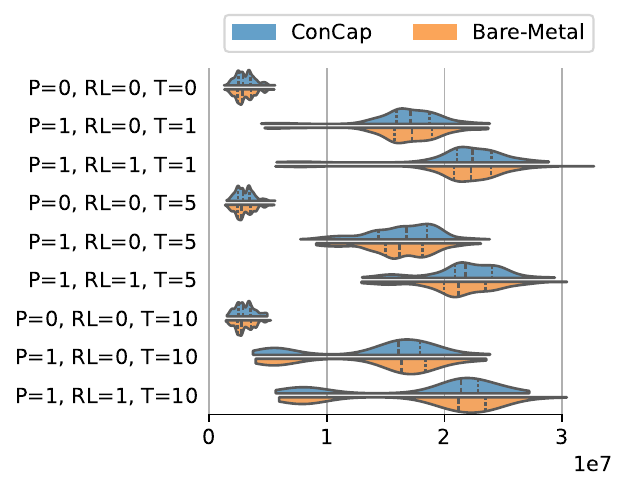}
        \caption*{Bwd IAT Total}
    \end{subfigure}
    \begin{subfigure}[b]{0.18\textwidth}
        \centering
        \includegraphics[width=\textwidth]{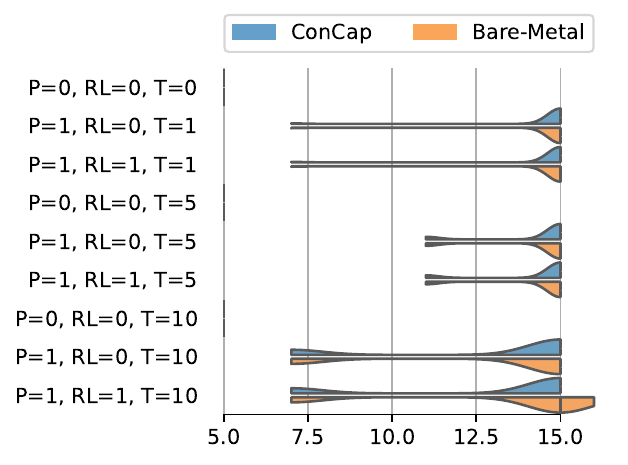}
        \caption*{Bwd PSH Flags}
    \end{subfigure}
    \begin{subfigure}[b]{0.18\textwidth}
        \centering
        \includegraphics[width=\textwidth]{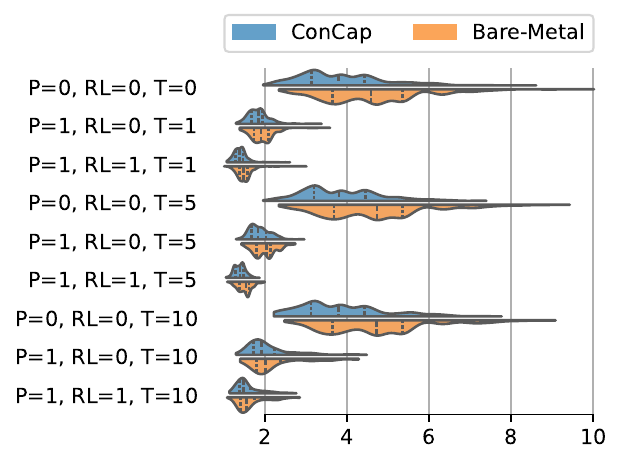}
        \caption*{Bwd Packets/s}
    \end{subfigure}
    \begin{subfigure}[b]{0.18\textwidth}
        \centering
        \includegraphics[width=\textwidth]{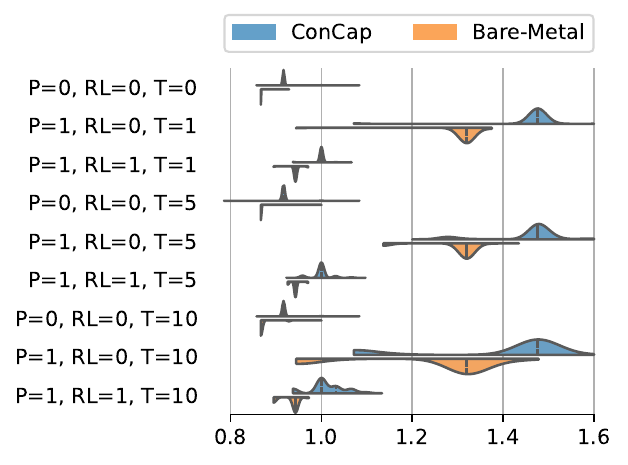}
        \caption*{Down/Up Ratio}
    \end{subfigure}
    \par\medskip
    \begin{subfigure}[b]{0.18\textwidth}
        \centering
        \includegraphics[width=\textwidth]{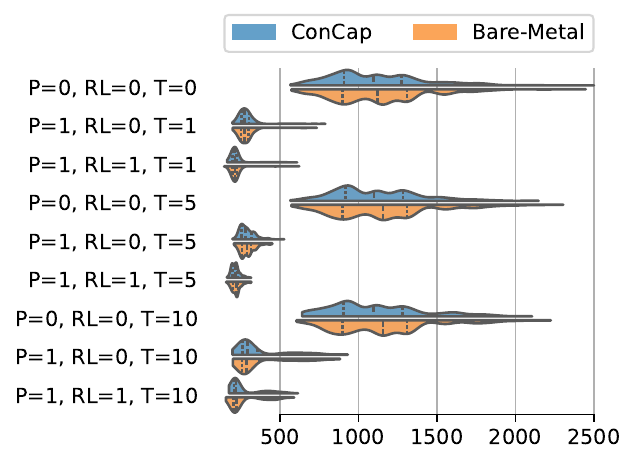}
        \caption*{Flow Bytes/s}
    \end{subfigure}
    \begin{subfigure}[b]{0.18\textwidth}
        \centering
        \includegraphics[width=\textwidth]{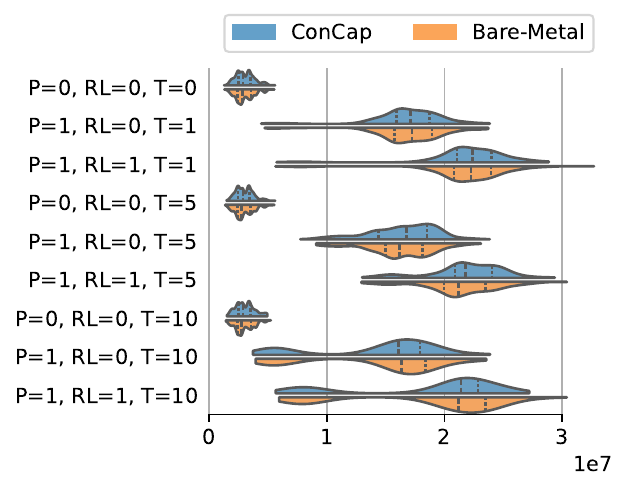}
        \caption*{Flow Duration}
    \end{subfigure}
    \begin{subfigure}[b]{0.18\textwidth}
        \centering
        \includegraphics[width=\textwidth]{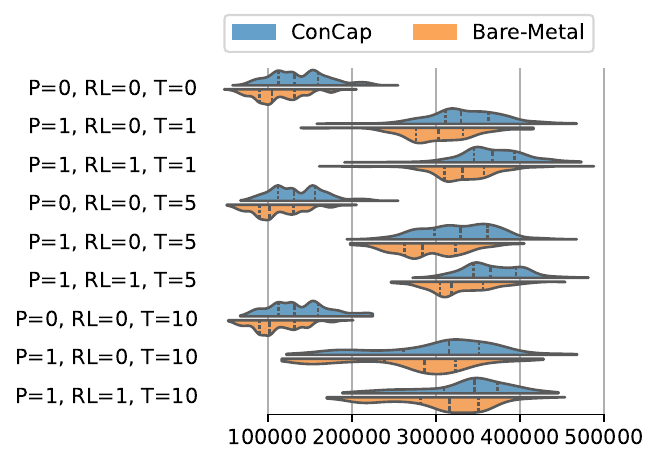}
        \caption*{Flow IAT Mean}
    \end{subfigure}
    \begin{subfigure}[b]{0.18\textwidth}
        \centering
        \includegraphics[width=\textwidth]{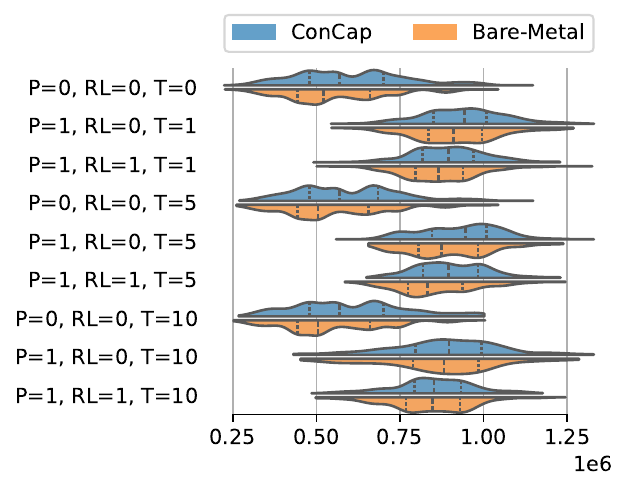}
        \caption*{Flow IAT Std}
    \end{subfigure}
    \begin{subfigure}[b]{0.18\textwidth}
        \centering
        \includegraphics[width=\textwidth]{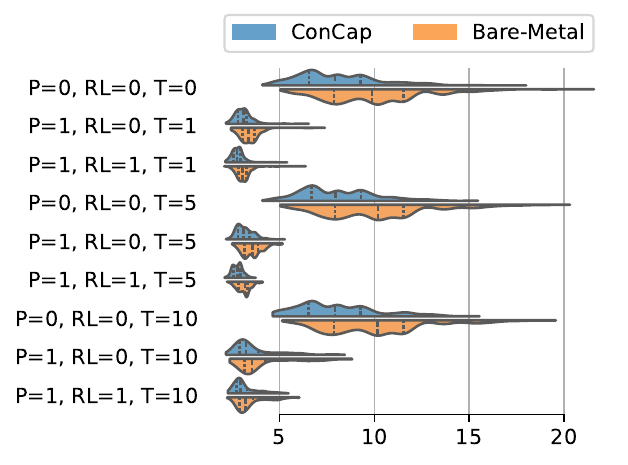}
        \caption*{Flow Packets/s}
    \end{subfigure}
    \par\medskip
    \begin{subfigure}[b]{0.18\textwidth}
        \centering
        \includegraphics[width=\textwidth]{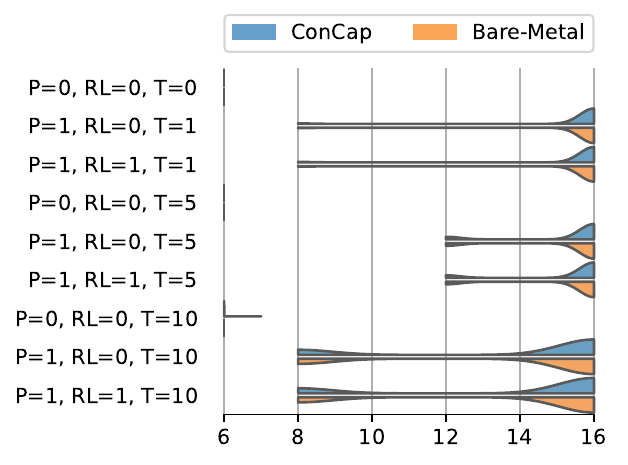}
        \caption*{Fwd Act Data Packets}
    \end{subfigure}
    \begin{subfigure}[b]{0.18\textwidth}
        \centering
        \includegraphics[width=\textwidth]{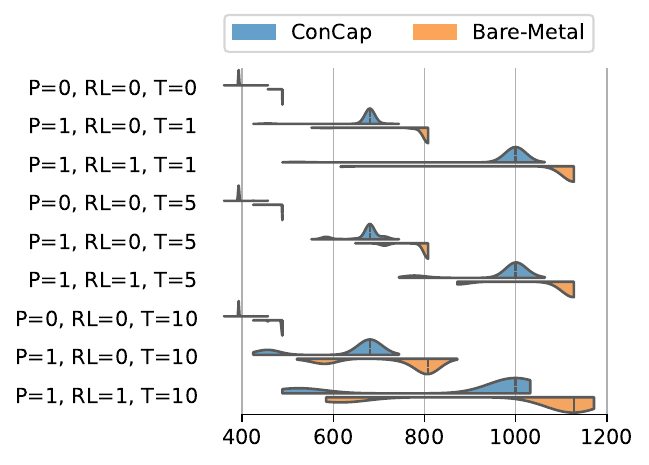}
        \caption*{Fwd Header Length}
    \end{subfigure}
    \begin{subfigure}[b]{0.18\textwidth}
        \centering
        \includegraphics[width=\textwidth]{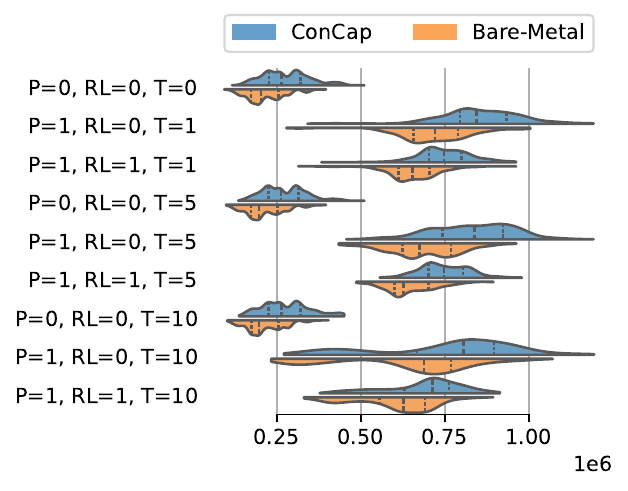}
        \caption*{Fwd IAT Mean}
    \end{subfigure}
    \begin{subfigure}[b]{0.18\textwidth}
        \centering
        \includegraphics[width=\textwidth]{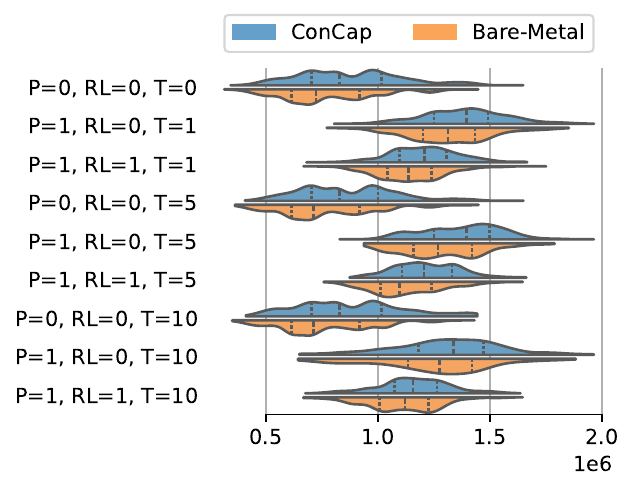}
        \caption*{Fwd IAT Std}
    \end{subfigure}
    \begin{subfigure}[b]{0.18\textwidth}
        \centering
        \includegraphics[width=\textwidth]{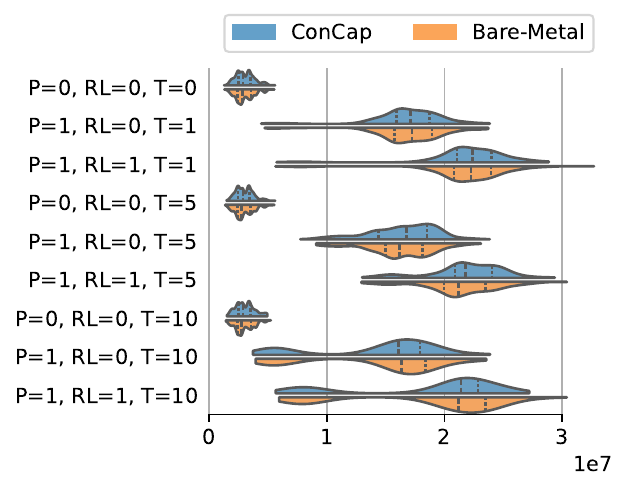}
        \caption*{Fwd IAT Total}
    \end{subfigure}
    \par\medskip
    \begin{subfigure}[b]{0.18\textwidth}
        \centering
        \includegraphics[width=\textwidth]{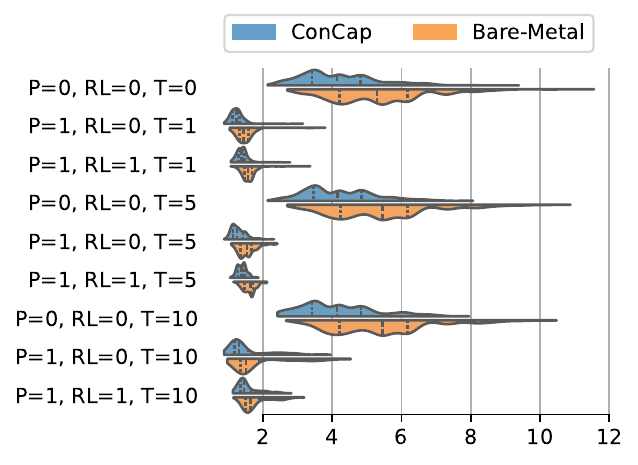}
        \caption*{Fwd Packets/s}
    \end{subfigure}
    \begin{subfigure}[b]{0.18\textwidth}
        \centering
        \includegraphics[width=\textwidth]{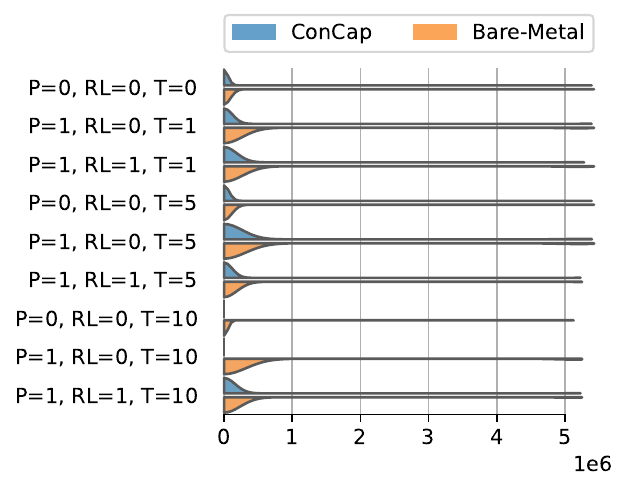}
        \caption*{Idle Mean}
    \end{subfigure}
    \begin{subfigure}[b]{0.18\textwidth}
        \centering
        \includegraphics[width=\textwidth]{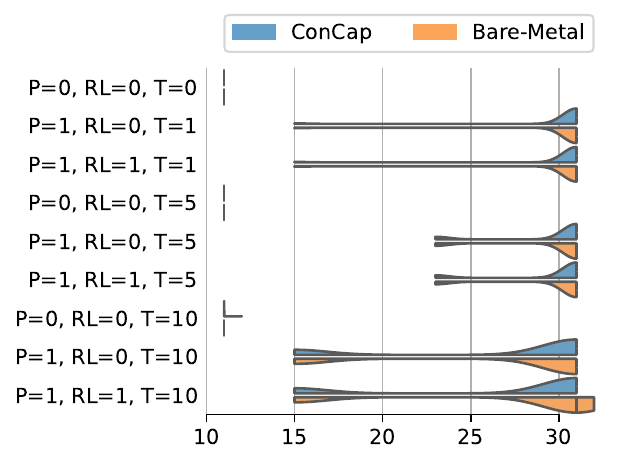}
        \caption*{PSH Flag Count}
    \end{subfigure}
    \begin{subfigure}[b]{0.18\textwidth}
        \centering
        \includegraphics[width=\textwidth]{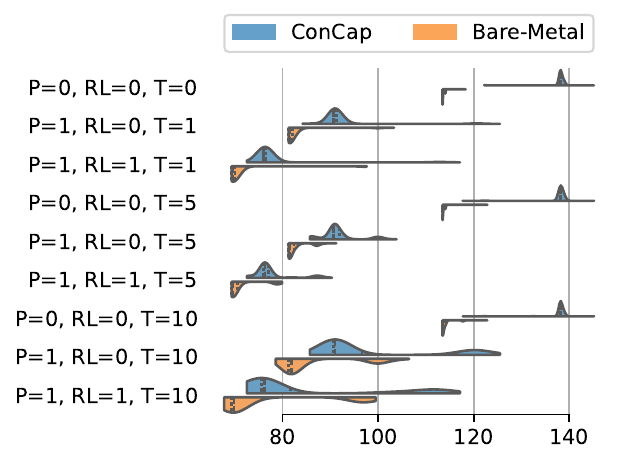}
        \caption*{Packet Length Mean}
    \end{subfigure}
    \begin{subfigure}[b]{0.18\textwidth}
        \centering
        \includegraphics[width=\textwidth]{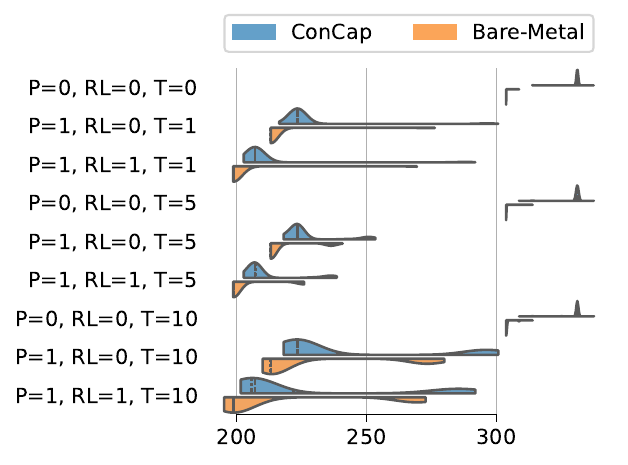}
        \caption*{Packet Length Std}
    \end{subfigure}
    \par\medskip
    \begin{subfigure}[b]{0.18\textwidth}
        \centering
        \includegraphics[width=\textwidth]{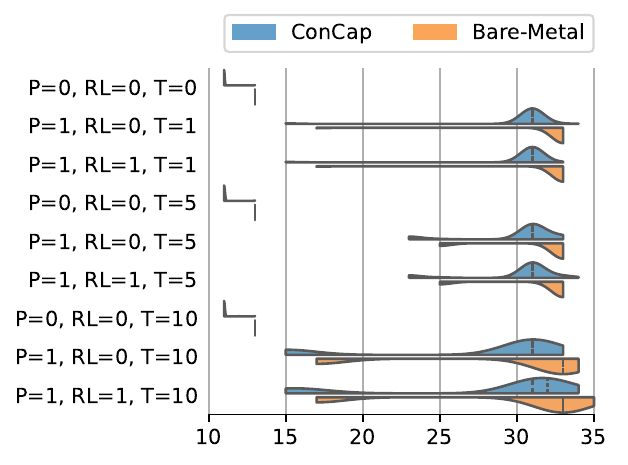}
        \caption*{Total Bwd Packets}
    \end{subfigure}
    \begin{subfigure}[b]{0.18\textwidth}
        \centering
        \includegraphics[width=\textwidth]{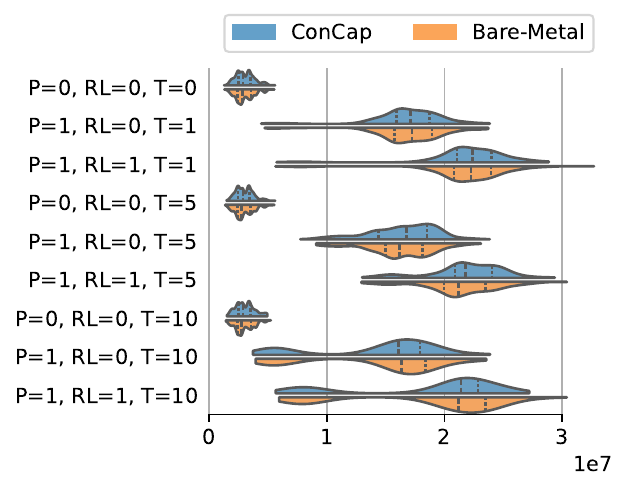}
        \caption*{Total Connection Flow Time}
    \end{subfigure}
    \begin{subfigure}[b]{0.18\textwidth}
        \centering
        \includegraphics[width=\textwidth]{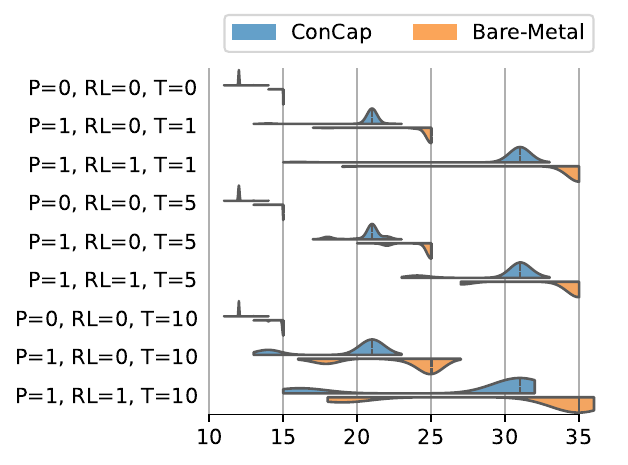}
        \caption*{Total Fwd Packet}
    \end{subfigure}
    \begin{subfigure}[b]{0.18\textwidth}
        \centering
        \includegraphics[width=\textwidth]{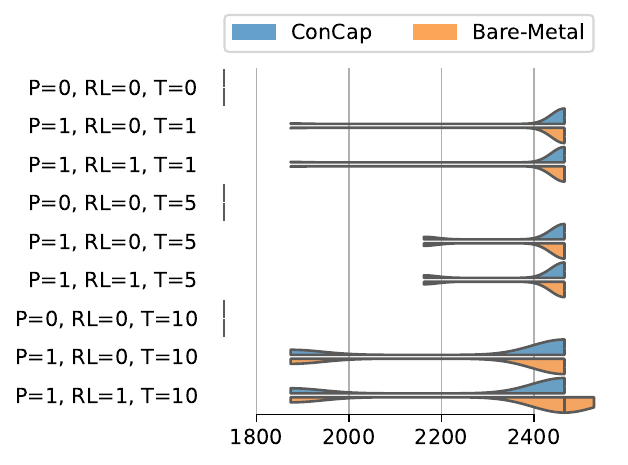}
        \caption*{Total Length of Bwd Packet}
    \end{subfigure}
    \begin{subfigure}[b]{0.18\textwidth}
        \centering
        \includegraphics[width=\textwidth]{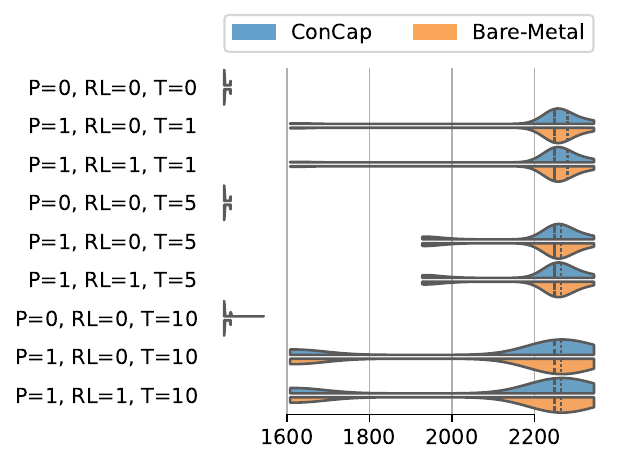}
        \caption*{Total Length of Fwd Packet}
    \end{subfigure}
    \caption{Comparison of NetFlow feature distributions of a Patator brute-force SSH attack between ConCap and a real network.}
    \label{fig:feature_dist}
\end{figure*}

\end{document}